\documentstyle[preprint,aps,graphicx,amssymb]{revtex}

\tightenlines

\newcommand{\etal}{{\em et al.}}                
\newcommand{\beq}{\begin{equation}}
\newcommand{\eeq}{\end{equation}}
\newcommand{\bea}{\begin{eqnarray}}
\newcommand{\eea}{\end{eqnarray}}
\newcommand{\beqars}[1]{\begin{eqnarray*}{#1}}
\newcommand{\eeqars}{\end{eqnarray*}}
\newcommand{\plb}[1]{Phys.~Lett.~B {#1}}
\newcommand{\zpa}[1]{Zeit.~f.~Phys.~{A#1}}

\newcommand{\npa}[1]{Nucl.~Phys.~{A#1}}

\newcommand{\pdd}{P.~Danielewicz}

\def\){\right) }
\def\({\left( }
\def\]{\right] }
\def\[{\left[ }

\begin{document}
\draft

\title{
\begin{flushright}
{\rm MSUCL-1143}
\end{flushright}
Determination of the
Mean-Field
Momentum-Dependence\\ using
Elliptic Flow
}

\author{
Pawe\l~Danielewicz\thanks{e-mail:
danielewicz@nscl.msu.edu}
}
\address{
National Superconducting Cyclotron Laboratory and\\
Department of Physics and Astronomy, Michigan State University,
\\
East Lansing, Michigan 48824, USA\\
}
\date{December 11, 1999}
\maketitle

\begin{abstract}
Midrapidity nucleon elliptic flow is studied within the
Boltzmann-equation simulations of symmetric heavy-ion
collisions.
The~simulations follow a~lattice Hamiltonian
extended to relativistic transport.  It is demonstrated
that in the peripheral heavy-ion collisions the~high-momentum
elliptic flow is strongly sensitive to the momentum dependence
of mean field at supranormal densities.  The~high
transverse-momentum particles are directly and exclusively
emitted from the high-density zone in the collisions, while
remaining particles primarily continue along the beam axis.
The~elliptic flow was measured by the KaoS Collaboration as
a~function of the transverse
momentum at a~number of impact parameters in Bi + Bi collisions
at 400, 700, and 1000~MeV/nucleon.
The~observed elliptic anisotropies in peripheral collisions,
which quickly rise with momentum, can only be explained in
simulations when assuming a~strong momentum dependence of
nucleonic mean field.  This momentum dependence must strengthen
with the rise of density above normal.  The~mean-field
parametrizations, which describe the data in simulations with
various success, are confronted with mean fields from
microscopic nuclear-matter calculations.  Two of the
microscopic
potentials in the comparisons have unacceptably weak
momentum-dependencies at supranormal densities.  The~optical
potentials from the Dirac-Brueckner-Hartree-Fock calculations,
on the other hand, together with the UV14 + TNI potential from
variational calculations, agree rather well within the region
of sensitivity with the parametrized potentials that best
describe the data.  \\[3ex]

\end{abstract}


\narrowtext

\section{Introduction}

Central collisions of heavy nuclei give, principally,
a~unique opportunity to access the properties of nuclear
matter at high
baryon densities and at high excitation energies.
In~practice, though, these collisions
often represent a~complex puzzle where many
effects, related to different matter properties, compete in
generating
observables.  If one takes e.g.\ the collective flow in
central collisions, it
is generally known to be affected by the
dependence of nuclear mean
field (MF) on baryon density and on particle momentum, as well
as by
the magnitude of in-medium nucleon cross-sections.  Different
sets of assumptions on the dependencies and on the cross
sections could be used to theoretically explain individual
data on flow.

For the sake of progress, under the
circumstances indicated above, it becomes of utmost
importance to isolate
either regions of data or data combinations that are
reasonably
sensitive to individual rather than many uncertain features of
nuclear
matter, simplifying the reaction puzzle.  Here, we concentrate
on the momentum dependence of the nucleonic optical potential.
The~momentum
dependence of that potential has been investigated at normal
and subnormal densities
utilizing~\cite{men71,sch82,ham90,kle94}  nucleon-nucleus
elastic scattering data
and different microscopic predictions have been
made~\cite{wir88,bal89,lim93,ins94,lee97}
with regard to the densities above normal.  Though supranormal
densities are reached in central heavy-ion reactions, only
circumstantial evidence for the momentum dependence of
the potential was seen, primarily in
flow~\cite{pan93,zha94,pak96,hun96},
and any quantitative assessment of that dependence was out of
question.  Here, we show that focussing on particular
features of the collective flow allows not only to see the
momentum dependence directly but, in fact, to assess it
quantitatively at the supranormal densities.

The momentum dependence posed difficulties for low-energy
central-reaction simulations due to problems with energy
conservation.
We elaborate on recent advances that expanded up and down the
beam energy range where simulations can be done reliably.

In the next section, we discuss at some length the framework of
our investigations.  In Sec.~3 we draw conclusions on the
optical potential by comparing results of the reaction
simulations with specific data on the elliptic flow.
In~Sec.~4 we confront the findings with the microscopic
calculations of the optical potential.  We summarize our results in
Sec.~5.

\section{Transport Theory}

\subsection{The Boltzmann Equation}

We follow the dynamics of central collisions of heavy nuclei
within the Landau quasiparticle theory of which the
relativistic formulation has been given in~\cite{bay76}.
In the quasiparticle approximation, the~state of a~system is
completely specified when the phase-space distributions $f_X
\equiv f_X ({\bf p}, {\bf r}, t)$
for all particles are given.  The~distributions satisfy the
Boltzmann equation
\begin{eqnarray}
{\partial f_X \over \partial t} + {\partial \epsilon_X \over
\partial {\bf p}} \, {\partial f_X \over \partial
{\bf r}} - {\partial \epsilon_X
\over \partial {\bf r}} \, {\partial f_X \over
\partial
{\bf p}} = {\cal K}^<_X \, (1 \mp f_X) - {\cal K}^>_X \, f_X \,
.
\label{boltz}
\end{eqnarray}
The l.h.s.\ accounts for the motion of particles in the MF,
while the r.h.s.\ accounts for collisions.
The~single-particle energies are derivatives of the net
energy~$E$ of a~system, with respect to the particle number:
\beq
\epsilon_X ({\bf p}, {\bf r}, t) = {(2 \pi)^3 \over g_X} \,
{\delta  E \over \delta f({\bf p}, {\bf r}, t)} \, ,
\label{epsilon}
\eeq
where $g_X$ is spin degeneracy.
Vector ${\bf v}_X = \partial \epsilon_X / \partial {\bf p}$
on the l.h.s.\ of~(\ref{boltz}) is velocity, while $- \partial
\epsilon_X / \partial {\bf r}$ is force.
The degrees of freedom in the description are nucleons, pions,
$\Delta$ and $N^*$ resonances, and, optionally, light ($A \le
3$) clusters.  Factors ${\cal K}^<$ and ${\cal K}^>$ on the
r.h.s.\ of~(\ref{boltz}) are the feeding and removal rates.

The net energy of a~system forms, obviously,
a~four-vector with the total momentum.  Since the
single-particle momentum~${\bf p}$
may be represented as a~derivative of the total momentum with
respect to~$f$, similarly to (\ref{epsilon}), and $f$ is
a~Lorentz scalar, it follows that $(\epsilon ({\bf p}), {\bf
p})$ form a four-vector.  This is independent of the specific
dependence of $\epsilon$ on ${\bf p}$ in any one frame.  By
considering a~change of the frame, one can subsequently see
that $\partial \epsilon / \partial
{\bf p}$ transforms as a~velocity in the Lorentz sense, so that
$(\gamma, \gamma \, {\bf v}) \equiv u$, with $\gamma =
1/\sqrt{1 - v^2}$, form a~four-vector.

The combination of relativity and momentum dependence brings in
some pecularities in the collision rates and cross sections,
beyond what is encountered nonrelativistically.  Thus,
consistently with the Fermi Golden Rule and the requirements
of covariance, the~contribution of binary
collisions of particles $X$ to the removal rate
in~(\ref{boltz}) is
\bea
\nonumber
{\cal K}^>_X ({\bf p_1}) & = & {g_X \over \gamma_1} \int {d{\bf
p}_2
\over (2  \pi)^3 \, \gamma_2} \int {d{\bf p}_1' \over (2
\pi)^3 \, \gamma_1'} \int {d{\bf p}_2' \over (2 \, \pi)^3 \,
\gamma_2'} \, {1 \over 2} \overline{| {\cal M}_{2X\rightarrow
2X'}|^2} \, \\ \nonumber && \times (2 \pi)^3 \, \delta({\bf
p}_1 + {\bf p}_2 -
{\bf p}_1' - {\bf p}_2') \, 2 \, \pi \, \delta
(\epsilon_1 + \epsilon_2
- \epsilon_1' - \epsilon_2') \, f_2 \, (1-f_1') \, (1-f_2')
\, \\ \nonumber & = &
{g_X \over \gamma_1} \int {d{\bf
p}_2 \over (2  \pi)^3 \, \gamma_2} \, {1 \over 2} \int  {d
\Omega^{*'} } \,  {p^{*'2} \over 4  \pi^2 \,
\gamma_1^{*'} \, \gamma_2^{*'}  \, v_{12}^{*'} } \,
 \overline{| {\cal M}_{2X\rightarrow
2X'}|^2} \, f_2 \, (1-f_1') \, (1-f_2')
\, \\ & = &
g_X \int {d{\bf
p}_2 \over (2  \pi)^3 } \, {1 \over 2} \int  {d
\Omega^{*'} } \,  v_{12} \, {d \sigma \over d \Omega^{*'}} \,
f_2 \, (1-f_1') \, (1-f_2')  \,  .
\label{rate}
\eea
In the above,
$\overline{| {\cal M} |^2}$ represents a~squared invariant
matrix element
for scattering, averaged over initial and summed over
final spin directions.  The~factors $\gamma$ are associated
with the respective velocities and
$d{\bf p}/\gamma$ is the invariant
measure as may be verified by considering the change of
a~reference frame.  The starred quantities in (\ref{rate})
refer to the two-particle c.m.\ defined by the vanishing of the
three-momentum, ${\bf P} = 0$, where $P = p_1 + p_2$.
The~cross section in~(\ref{rate}) is given by
\beq
{d \sigma \over d \Omega^{*'}} =
 {p^{*'2} \over 4 \pi^2 \,
\gamma_1^{*} \, \gamma_2^{*}  \, v_{12}^{*}  \,
\gamma_1^{*'} \, \gamma_2^{*'}  \, v_{12}^{*'} } \,
 \overline{| {\cal M}_{2X\rightarrow
2X'}|^2} \, .
\label{cross}
\eeq
The relativistic relative velocity in (\ref{rate}) and
(\ref{cross}) is defined through
\beq
\gamma_1 \, \gamma_2 \, v_{12} = \left[ - {[ (P\, u_2) \, u_1 -
(P \, u_1 ) \, u_2 ]^2 \over P^2 } \right]^{1/2} \, .
\eeq
The~above definitions ensure the standard form of the detailed
balance relation, i.e.\ here
\beq
p^{*2} \, {d \sigma \over d \Omega^{*'}}
=
p^{*'2} \, {d \sigma \over d \Omega^{*}} \, .
\eeq
In the c.m., the relative velocity reduces to the velocity
difference.  The factor of $1/2$ in front of the
angular integrations in (\ref{rate}) accounts, in the standard
manner, for
the double-counting of the final states in scattering for like
particles.

\subsection{The Energy Functional}

The MF dynamics follows from the dependence of
total energy on the phase-space distributions.
We adopt simple parametrizations for the net energy,
that permit the transport calculations to be carried through,
and that are flexible enough with regard to the MF
and to the equation of state~(EOS).  In our
parametrizations,
the energy consists of the covariant volume term and of the
noncovariant gradient-correction, isospin interaction, and
Coulomb terms defined in the system frame:
\beq
E = \int d{\bf r} \, \tilde{e} + E_1 + E_T + E_{coul} \, .
\label{E=}
\eeq
In the practice of central
reactions, the importance of
covariance and the indispensability of the correction terms are
usually mutually exclusive.
For different systems, there is a~level of cancellation
between the Coulomb and isospin terms.
The~Coulomb energy in (\ref{E=}) is simply
\beq
E_{coul} = {1 \over 4 \pi  \epsilon_0} \int d{\bf r} \, d
{\bf r}' \, {\rho_{ch} ({\bf
r}) \, \rho_{ch} ({\bf r}') \over |{\bf r} - {\bf r}'| } \, .
\eeq
The~gradient term is
\beq
E_1 = {a_1 \over 2  \rho_0} \int d{\bf r} \, (\nabla \rho)^2 \,
,
\label{E1}
\eeq
where $\rho$ denotes the baryon density and $\rho_0
=0.160$~fm$^{-3}$
is the normal density.  Finally, the isospin interaction term
is
\beq
E_T = {a_T \over 2  \rho_0} \int d{\bf r} \, \rho_T^2 \, ,
\label{isospin}
\eeq
where $\rho_T$ represents the density of the third component of
isospin.

The gradient term in the energy allows us, primarily, to
account for
the effect of the finite range of nuclear forces, augmanted by
the lowest-order quantal effect of the curvature in the
wavefunctions, in the Thomas-Fermi (TF)
initialization of nuclei~\cite{hol77,len89} for our reaction
simulations.  We take the coefficient in (\ref{E1})
equal to
$a_1 = 21.4$~MeV~fm$^2$
for momentum-independent~MFs,
as corresponding to the finite-range correction from the Skyrme
effective interaction and an addition
from the Weizs\"{a}cker kinetic-energy term~\cite{rin80}.  For
the momentum-dependent fields, we take a~bit lower $a_1
= 18.2$~MeV~fm$^2$ from our own adjustments to ground-state
data. The~initialization is described in the next
subsection.

The isospin term~(\ref{isospin}) in the energy contributes
to the isospin asymmetry
coefficient in the Weizs\"{a}cker mass formula (at a~50\%
level) and gives rise
to the isospin term in the optical potential, of the magnitude
${a_T \over 4} \, {|N - Z| \over A}$ for nucleon scattering off
a~nucleus.  We use $a_T = 97$~MeV,
that simultaneously produces reasonable results for the mass
asymmetry and the asymmetry in the potential deduced from
nucleon scattering~\cite{bec69,sch82}.

Aside from the above isospin component, the strong-interaction
field (from the covariant volume part) is chosen as acting only
on baryons in our calculations.  That is the field that we are
interested
in here.  Pions are, anyway, infrequent in the energy range
within which
we will be making comparisons to data.  We should note that,
when the vector
and scalar type~MFs may be momentum dependent with no
exclusive
dependence on the vector and scalar densities, there is neither
a~benefit nor a~phenomenological basis, in the absence of spin
dynamics, for a~separate consideration of these fields.

Guided solely by the calculational convenience, we choose
the fields that could be easily identified as either vector or
scalar~\cite{pan93}.  Thus, in the case of the fields
{\em without momentum dependence} in their nonrelativistic
reduction, we use the energy density in the form~\cite{dan98}
\beq
\tilde{e} = \sum_X {g_X } \int {d{\bf p} \over (2 \pi)^3 } \,
f_X ({\bf p}) \, \sqrt{p^2 + m_X^2 (\rho_s)}
+ \int_0^{\rho_s} d{\rho_s}' \, U(\rho_s') -
\rho_s \, U(\rho_s),
\label{e=}
\eeq
where~$m_X (\rho_s) = m_X + A_X \,
U(\rho_s)$, $A_X$~is~baryon~number,~and
\beq
\rho_s = \sum_X
{g_X} \, A_X \int
{d{\bf p} \over (2 \pi)^3 } \, {m_X (\rho_s) \over \sqrt{p^2 +
m_X^2 (\rho_s)}}
\, f_X ({\bf p}) \, .
\eeq
The~energy~(\ref{e=}) alone gives rise to
single-particle energies
\beq
\tilde{\epsilon}_X (p, \rho_s) = \sqrt{p^2 + m_X^2(\rho_s)}.
\label{tepsilon}
\eeq
We take
\beq
U({ { \xi}}) = {-a \, {{\xi}} + b \, {{\xi }}^\nu \over 1 +
({{ \xi }}/2.5)^{\nu - 1}} \, ,
\label{U=}
\eeq
with ${{\xi }} = \rho_s / \rho_0$, and $a$, $b$,
and $\nu$ adjusted to produce average nuclear ground-state
properties.
The~role of the denominator in (\ref{U=}) is to prevent
supraluminous behavior at high densities.
To the energies (\ref{tepsilon}), we add in the system frame
the gradient, isospin, and Coulomb corrections, that
contribute to the forces, but drop out from collision
integrals and velocities,
\beq
\epsilon_X = \tilde{\epsilon}_X + A_X \, U_1 + t_{3X} \, U_T +
Z_X \, \Phi \, ,
\label{eX}
\eeq
where $U_1 = - a_1 \, \nabla^2 (\rho/\rho_0)$, $U_T = a_T \,
\rho_T/\rho_0$, and $\Phi$ is the Coulomb
potential.

To determine $a$, $b$, and $\nu$, we
required the energy per nucleon
to minimize in nuclear matter at $\rho = \rho_0$ at the value
of $e/\rho - m_N \approx - 16.0$~MeV for incompressibility $K =
210$~MeV, and at $- 17.0$~MeV for $K = 380$~MeV.  For the
higher~$K$, the~energetic cost for the surface is
higher.  That leads to difficulties, for the
TF theory, in reproducing the~average dependence of nuclear
binding energy on mass
number (especially in the low mass region), which we
partly compensate for with the stronger binding in the
infinite-volume
limit.\footnote{
A~more thorough discussion can be
found
in~\protect\cite{mye98};
adjustments of $a_1$ in~(\ref{E1}) cannot be done
without worsening the TF description of measured rms radii.
The~fact that we employ the average of proton and neutron
masses for nucleons is
of no importance for the issue.}
The~parameter sets resulting from adjustments
are: $a=187.24$~MeV, $b=102.62$~MeV, and~$\nu=1.6339$ for
$K=210$~MeV, and $a=121.26$~MeV, $b=52.10$~MeV, and
$\nu=2.4624$ for $K=380$~MeV.  Generally,
the~reproducing of the binding-energy curve matters for
assessing the excitation energy of transients formed in
low-energy or peripheral reactions.

        In the case of MFs {\em dependent on momentum} in their
nonrelativistic reduction, we parametrize the energy density in
the local frame where baryon flux vanishes~\cite{pan93,dan98}
(see also~\cite{hom99}), ${\bf J} = \sum_X g_X \, A_X
\, \int {d{\bf p} \, \over (2 \pi)^3} f_X \,
{\bf v}_X = 0$, with
\begin{equation}
\tilde{e} = \sum_X {g_X } \int {d{\bf p} \over (2 \pi)^3 } \,
f_X ({\bf
p}) \left( m_X + \int_0^p dp' \, v_X^*(p', \rho) \right) +
\int_0^\rho d\rho ' \, U(\rho '),
\label{emo=}
\end{equation}
where  $U$ is of the form expressed by Eq.~(\ref{U=}), with
$\xi = \rho/\rho_0$, and the
local particle velocity $v_X^*$ depends on (kinematic) momentum
and density through
\begin{equation}
v_X^*(p,{{ \xi }}) = {p \over \sqrt{p^2 + m_X^2
\left/ \left( 1 + c \, {m_N \over m_X}  \, {{{ A_X \, \xi }}
\over (1 + \lambda \, p^2/m_X^2)^2} \right)^2 \right. } } \, .
\label{vX}
\end{equation}
The energy (\ref{emo=}) alone yields the local single-particle
energies
\begin{equation}
\tilde{\epsilon}_X(p, \rho) = m_X + \int_0^p dp' \, v_X^* + A_X
\left[  \rho
\left\langle \int_0^{p_1} dp' \, {\partial v \over \partial
\rho} \right\rangle + U(\rho) \right] \, ,
\label{eps=}
\end{equation}
where
\beq
\rho \left\langle \int_0^{p_1} dp' \, {\partial v \over
\partial \rho}
\right\rangle = \sum_Y g_Y \int {d {\bf p}_1 \over (2 \pi)^3}
\, f_Y ({\bf p}_1) \int_0^{p_1} dp' \, {\partial v_Y^* \over
\partial \rho} \, .
\eeq
In their nonrelativistic reduction, the energies~(\ref{eps=})
are similar to the energies proposed for the nonrelativistic
transport by Bertsch, Das Gupta {\em et
al.}~\cite{ber88,cse92}.  Principally, for matter without
local reflection symmetry in momentum space, there is
a~correction term
in Eq.~(\ref{eps=}) from the condition ${\bf J} = 0$, which
we ignore as it would seriously complicate our
calculation.  Suprisingly,
that term is never mentioned in the context of nonrelativistic
calculations.  On~the other hand, in practice, the~omission of
that term causes no problems with the energy-momentum
conservation
of any relevance, as will be indicated.  Different sets of
parameters for~(\ref{emo=}), giving different values for the
group effective mass~\cite{jam89} in normal matter
at the Fermi surface ($m^*= p^F/v^F$) and for the
incompressibility, are exhibited in Table~\ref{tabmo}.

\subsection{The Thomas-Fermi Equations}

The requirement that the energy (\ref{E=}) is minimal in the
ground state, for
a~nucleus with a~definite number of protons and neutrons,
yields the set of TF equations:
\bea
\label{TFp}
0 & = & \tilde{\epsilon}_p \left( p^F(\rho_p) \right) - a_1 \,
\nabla^2 \left( {\rho \over \rho_0} \right) + {a_T \over 4} \,
{\rho_p - \rho_n \over \rho_0} + \Phi - \mu_p \, , \\
0 & = & \tilde{\epsilon}_n \left( p^F(\rho_n) \right) - a_1 \,
\nabla^2 \left( {\rho \over \rho_0} \right) - {a_T \over 4} \,
{\rho_p - \rho_n \over \rho_0} - \mu_n \, ,
\label{TFn}
\eea
and the condition $\nabla_n \rho = 0$ at the edge of the
density distribution.  In~the equations, $\mu_p$ and $\mu_n$
are the Lagrange multipliers for the proton and neutron
numbers,
respectively.

The~role of the derivative correction in
(\ref{TFp}), (\ref{TFn}), and (\ref{eX}) is to reduce the
effect of the negative MF when the density distribution
in the vicinity is primarily concave and to enhance the
effect of the field when the density is convex.  Such a~result
would be obtained for a~finite-range
effective two-body interaction convoluted with density expanded
in position to second order.  Not surprisingly, the derivative
correction is small
but it becomes important when the energies balance, permitting
an~adequate description of the density in the ground
state.  For nucleons with high momenta relative to a~system,
the sign of the correction, obviously, becomes an issue.

In finding the density profile, it is convenient
to transform the~TF equations into:
\bea
\label{drho}
&&{1 \over r^2} \, {d \over dr} \, r^2 \, {d \over dr} \, \rho
 =  {\rho_0 \over 2 a_1 } \left[ \tilde{\epsilon}_p^F +
 \Phi + \tilde{\epsilon}_n^F - \mu_p - \mu_n \right] \, ,
\\
&&\mu_p - \mu_n =  \tilde{\epsilon}_p^F + \Phi -
\tilde{\epsilon}_n^F + {a_T \over 2} \, {\rho_p - \rho_n \over
\rho_0} \, .
\label{mupn}
\eea
The net density profile $\rho (r)$ may be obtained by starting
Eq.~(\ref{drho}) with some density at $r=0$.  At any $r$,
separate $\rho_p$ and $\rho_n$ may be found from~(\ref{mupn})
and $\Phi$ can be obtained from the Gauss' law.  The~acceptable
starting $\rho(r=0)$ is the one for which $\rho=0$ is reached
in the solution simultaneously with $d \rho /dr = 0$.
The~chemical potentials are adjusted until the required proton
and neutron numbers are obtained.  At the end, $\Phi$ and
$\mu_p$ may be renormalized, so that $\Phi \rightarrow 0$ as $r
\rightarrow \infty$.

\begin{figure}

\centerline{\includegraphics[angle=180,
width=.70\linewidth]{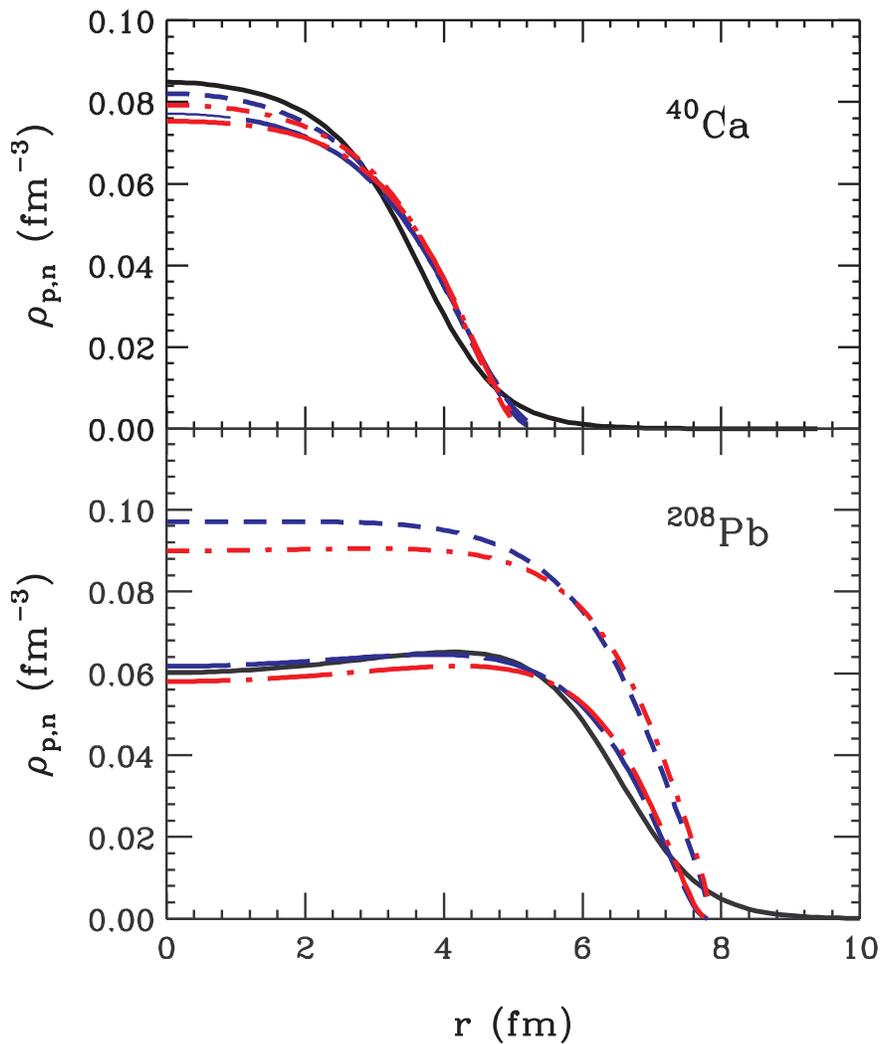}}
\vspace*{.1in}
\caption{
Nucleon density profiles from solving the TF equations for
MFs corresponding to $K=210$~MeV, together with the
empirical charge density profiles for $^{40}$Ca and $^{208}$Pb.
The~solid lines represent the empirical profiles from
Ref.~\protect\cite{jag74}.  The~long- and short-dashed lines
represent the proton and neutron profiles, respectively, for
the momentum-independent field.  The~long- and
short-dash-dotted
lines represent the proton and neutron profiles, respectively,
for the momentum-dependent field in Table~\protect\ref{tabmo}
that yields $m^*=0.70\, m$.
}
\label{proden}
\end{figure}

Figure \ref{proden} shows the calculated proton and neutron
density profiles for a~moderate and large nucleus and
MFs corresponding to $K=210$~MeV, with and without
momentum-dependence.  Also the empirical charge densities are
shown.

\subsection{Optical Potential}

Comparing
relativistic MFs, e.g.\ when employing both scalar
and vector potentials with momentum dependence,
and nonrelativistic fields,
poses some difficulty.  Wave approaches are
summarized at times in terms of the Schr\"{o}dinger equivalent
potential.  Feldmeier and Lindner~\cite{fel91} (but see
also~\cite{har87}) suggested the
definition of the optical potential in nuclear matter as simply
the difference:
\beq
\label{Uopt}
U^{opt}(p) = \epsilon(p) - m - T(p) \, ,
\eeq
where $T$ is kinetic energy at the same momentum~$p$ in free
space.
The momentum can be expressed in terms of net energy (such as
that of incident nucleon), whether in a~relativistic or
nonrelativistic approach,
yielding~$U^{opt}(\epsilon)$.  The~resulting potential is
similar to the Schr\"{o}dinger-equivalent potential at low to
moderate momenta and has a~more satisfactory high-energy
limit.

\begin{figure}

\centerline{\includegraphics[angle=0,
width=.90\linewidth]{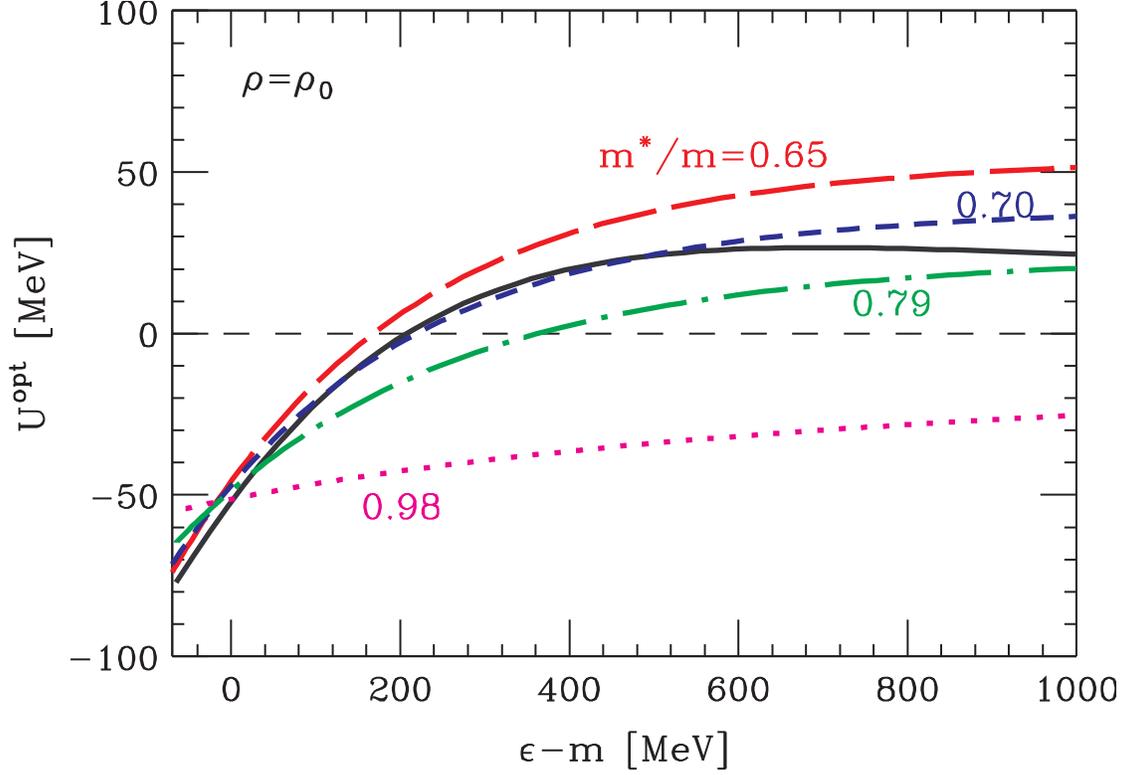}}
\vspace*{.1in}
\caption{
Real part of the optical potential in normal nuclear matter, as
a~function of nucleon energy.  The~solid
line represents a~fit~\protect\cite{fel91} to the information
from structure and nucleon
scattering~\protect\cite{per76,ham90}.  The~dashed lines
represent the potential for different MF
parametrizations in Table~\protect\ref{tabmo} for $K=210$~MeV.
The~numbers in the figure indicate the ratio~$m^*/m$.  For
$m^*/m = 0.65$ only the first set from the table is
represented.
The~dotted line represents the potential for the MF
without momentum dependence in its nonrelativistic reduction,
corresponding to $K=210$~MeV.
}
\label{upo1}
\end{figure}

Feldmeier and Lindner used (\ref{Uopt}) to represent the
nonrelativistic results in the work by Perey and
Perey~\cite{per76} in a~combination with the results of
an~analysis
of proton scattering data, in terms of relativistic MF
potentials, by Hama \etal \cite{ham90}.
The~best fit~\cite{fel91} with an analytic ansatz is shown by
a~solid line in Fig.~\ref{upo1}.  The~dashed lines in the
figure show potentials in normal matter for three
of our $K=210$~MeV parametrizations from Table~\ref{tabmo}.
The~omitted $K=380$~MeV and
$m^*/m=0.70$ potential is practically indistinguishable from
the displayed $K=210$~MeV and $m^*/m=0.70$ potential.
This is because the dependence of the {\em velocity} on
momentum
and density is forced to be the same for the two potentials,
cf.~Eq.~(\ref{vX}) and Table~\ref{tabmo}.
In Fig.~\ref{upo1},
the~linear dependence of the potentials on energy in
the low-energy
region indicates that the effective mass provides a~good
characterization of those potentials in that region.
The~use of scalar (rather than vector) density in MFs
without a~momentum
dependence in the nonrelativistic reduction gives rise to
a~weak dependence ($m^*/m = 0.98$ at
$\epsilon^F$) for relativistic energy, as illustrated by the
dotted line in Fig.~\ref{upo1}.

\begin{figure}

\centerline{\includegraphics[angle=0,
width=.70\linewidth]{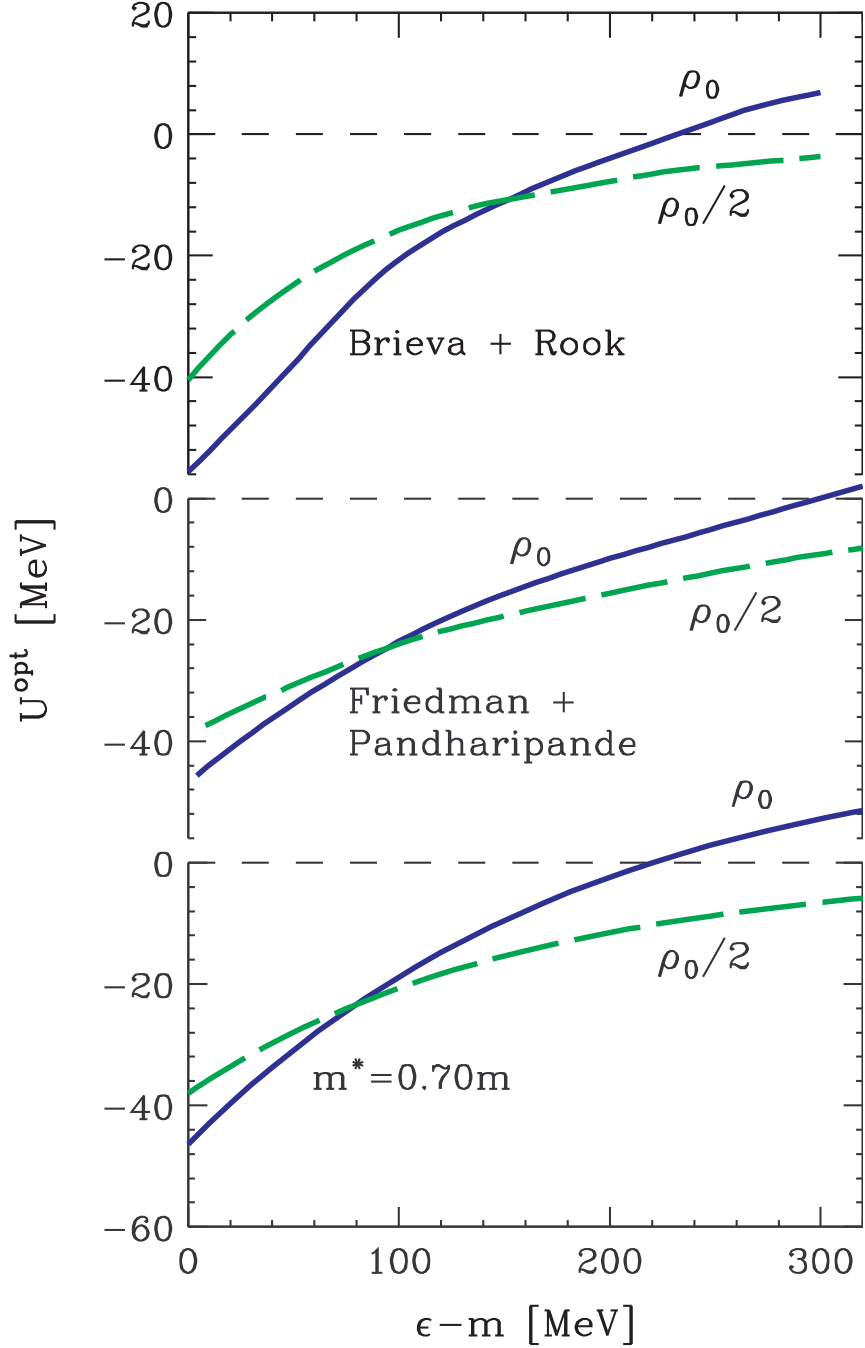}}
\vspace*{.1in}
\caption{
Optical potential in nuclear matter, as
a~function of nucleon energy, obtained in
Brueckner-Hartee-Fock~\protect\cite{bri77} (top) and
variational~\protect\cite{fri81} calculations and in our
$K=210$~MeV and
$m^*/m = 0.70$ parametrization.  Solid and dashed lines show
the potential at $\rho_0$ and $\rho_0/2$, respectively.
}
\label{upoden}
\end{figure}

With regard to other than normal densities, in some analysis of
high-energy scattering, see e.g.~\cite{sch82}, it was
found that potentials of a~bottle shape described the data
better
than standard Woods-Saxon potentials, with the interior at
times repulsive while the surface region attractive.  (That
was,
in fact, obtained in the analysis~\cite{ham90} of Hama \etal\
but, primarily, as a~byproduct
since the parametrization did not have enough flexibility to
address the surface directly.)  Such results indicate
differences in the momentum dependence at different densities,
with the potential changing sign at different energies at the
different densities.  That contradicts a~naive expectation that
the potential at a~lower density may be obtained by rescaling
of the potential at the normal density; the result is actually
predicted by nuclear-matter calculations.  Figure~\ref{upoden}
displays the potentials from early
Brueckner-Hartree-Fock~\cite{bri77} and
variational~\cite{fri81} calculations at
$\rho_0$ and $\rho_0/2$.  It~is seen that the $\rho_0/2$
results continue to be attractive up to higher energies than
the $\rho_0$ results.  That is also present in our
parametrizations.
The~increasing slope of a~potential with
energy, as the density rises, generally indicates
the increasing velocities with the rise in density, on account
of the momentum dependence.

The flow data, that we shall analyze, will turn out to be
sensitive to the potential at supranormal densities.
After the analysis, we~will compare our results
to the
microscopic calculations done at such densities.
Feldmeier and Lindner attribute an error of the order of 5~MeV
to~their optical potential fit.
In our analysis, we will try not to be guided directly by the
nucleon
data, but rather explore what the flow data alone may tell us about the
potential.

\subsection{Lattice Hamiltonian}

Advancing of physical conclusions from comparisons to data
depends on advances in the methods.
In solving the Boltzmann equation~(\ref{boltz}),
the~phase-space density is
represented in terms of a set of $\delta$-functions, or
test-particles,
\beq
{g \over (2\pi)^3} \, f ({\bf p}, {\bf r}, t )
= {1 \over {\cal N}} \sum_k \delta( {\bf p} - {\bf p}_k (t)) \,
\delta({\bf r} - {\bf r}_k (t)) \, ,
\eeq
where ${\cal N}$ is the number of test-particles per particle.
Such a representation is possible when, in its use, the
phase-space density gets integrated over
phase-space volumes incorporating many test-particles.  In the
solution of~(\ref{boltz}), the
changes in the density with time due to collisions are
accounted for
by a~Monte-Carlo procedure, while the changes described by the
l.h.s.\ of~(\ref{boltz}) are accounted for by requiring that
the test-particles follow the Hamilton's equations:
\bea
{d \, {\bf r}_k \over dt} & = & {d \, \epsilon \over d \, {\bf
p}_k} \, ,\\
{d \, {\bf p}_k \over dt} & = & - {d \, \epsilon \over d \, {\bf
r}_k} \, .
\eea

For the Hamilton's equations,
an~averaging in space must be done to obtain
single-particle energies and gradients must be calculated.
Often, a~mesh
in space is employed.  The~representation of the phase-space
distribution in terms of the test-particles leads to
statistical fluctuations in the driving terms in the Hamilton's
equations.  Due to these fluctuations the net energy and
net momentum of the system fluctuate.
As a~convex function of momenta, the~energy grows with
time as an~effect of the diffusive process.
The~problem is more serious for
momentum-dependent than momentum-independent MFs
because of stronger fluctuations.
The~former fields depend both on position and momentum of
the test-particles
and, additionally, the~driving terms fluctuate in two
Hamilton's equations rather than only in one.  The~problem may
be
so severe that, for momentum-dependent fields and for {\em ad
hoc} methods of calculating the
single-particle energies and gradients,
in a~slow central
reaction at few tens of MeV/nucleon, the~spurious energy gain
may get close to the net excitation energy.
Another example when fluctuations can be very harmful, is when
using the MF to simulate the chiral phase transition.
If such transition is not simulated though, the~situation
generally quite improves with the rise in the energy of
a~simulated reaction.

The fluctuations may be reduced by increasing the number of
test-particles~${\cal N}$, for the sake of improving the
accuracy of
a~calculation, but the inverse square-root covergence is slow.
That is worsened by the fact that, otherwise, an~improvement
in the accuracy requires
the~reduction in the spatial region for averaging and
gradients.
Thus, other ways of reducing fluctuations and their
consequences and of accelerating the convergence of
calculations
must be sought, on top of using a~large~${\cal N}$.
The~lattice Hamiltonian method proposed first by Lenk and
Pandharipande for momentum-independent fields~\cite{len89}
accomplishes three things.  The~method ensures a~continuous
variation with position of the test-particle contributions to
local densities, a~continuous variation of the test-particle
momenta with time,
and it imposes a~net energy constraint on the
dynamics.  The~constraint
plays a~local role, on the scale of the discretization in
space; this is achieved through a~consistency between the
computation of the test-particle contributions to the densities
and of the forces on the particles.
As a~rule, when following the method, an~excellant
momentum conservation also is observed for isolated fragments.
Here, we generalize the method to the
momentum-dependent and relativistic fields.
The~generalized method was
already utilized in~\cite{dan98} and was in parallel developed
in~\cite{per99}.

Within the computational region, we introduce a~mesh with nodes
at ${\bf r}_\alpha$ separated by $\Delta l_i$, $i=1,2,3$, in
three carthesian directions.   With each of the nodes, we
associate a~form factor~$0 \le S_\alpha \le 1$, continuous and
piecewise differentiable within
the computational region and concentrated around~${\bf
r}_\alpha$.  We require the form factors to satisfy:
\beq
\sum_\alpha S_\alpha ({\bf r} - {\bf r}_\alpha) = 1 \, ,
\eeq
for all ${\bf r}$ within the region, and
\beq
\int d{\bf r} \, S_\alpha ({\bf r}) = \Delta V \, ,
\eeq
where $\Delta V= \Delta l_1 \, \Delta l_2 \, \Delta l_3$.
The~second requirement is that, for averaging, every node gets
its share of the volume and the first requirement is that every
test-particle within the region is fully accounted for.
The~average Wigner function for a~node, in terms of $S_\alpha$,
is
\beq
\overline{f} ( {\bf p}, {\bf r}_\alpha ) = {1 \over {\cal N}
\, \Delta V}
\sum_k \delta({\bf p} -
{\bf p}_k) \, S_\alpha({\bf r}_k -{\bf r}_\alpha) \, .
\eeq
The~approximate energy (lattice Hamiltonian) in
terms of the spatially averaged Wigner functions is then
\bea
\nonumber
\overline{E} & = & \Delta V \, \sum_\alpha \Bigg(
\tilde{e}\lbrace
\overline{f}({\bf r}_\alpha ) \rbrace + {a_T \over 2 \rho_0} \,
\overline{\rho}_T^2
+ {1\over 2e} \, \overline{\rho}_{ch}({\bf
r}_\alpha) \, \overline{\Phi}({\bf r}_\alpha)\\
& & +
{a_1 \over 4 \rho_0} \sum_i {1 \over
(\Delta l_i)^2} \, \left[
\left( \overline{\rho}({\bf r}_\alpha + \Delta l_i
\, \hat{\bf u}_i) - \overline{\rho}({\bf r}_\alpha)
\right)^2
 +
\left( \overline{\rho}({\bf r}_\alpha)
- \overline{\rho}({\bf r}_\alpha - \Delta l_i \, \hat{\bf u}_i)
\right)^2
\right]  \Bigg)
\, ,
\eea
where $\overline{\Phi}$ satisfies a~discretized Poisson
equation~\cite{fel92} with~$\overline{\rho}$.

The single-particle energy for a~test-particle is the variation
of~$\overline{E}$ with respect to the particle number.
The~energy turns
out to be a~weighted average of energies associated with the
nodes in the vicinity:
\beq
\label{epnu}
\epsilon ({\bf p} ,{\bf r} ) = \sum_\alpha S_\alpha({\bf
r} - {\bf r}_\alpha) \, \overline{\epsilon}( {\bf p} , {\bf
r}_\alpha )  \, ,
\eeq
with
\beq
\overline{\epsilon}_X ({\bf p}, {\bf r}_\alpha) =
\tilde{\epsilon}_X ({\bf p}, \lbrace \overline{f} ({\bf
r}_\alpha) \rbrace) + A_X \, \overline{U}_1 ({\bf r}_\alpha) +
t_{3X} \, \overline{U}_T ({\bf r}_\alpha) + Z_X \,
\overline{\Phi} ({\bf r}_\alpha) \, ,
\eeq
where $\overline{U}_T ({\bf r}_\alpha) = a_T \,
\overline{\rho}_T ({\bf r}_\alpha)/\rho_0$ and
\beq
\nonumber
\overline{U}_1 ({\bf r}_\alpha) = {a_1 \over \rho_0} \sum_i {1
\over (\Delta l_i)^2} \,
\Big(
2\overline{\rho}({\bf r}_\alpha)-
\overline{\rho}({\bf r}_\alpha + \Delta l_i
\, \hat{\bf u}_i) -
\overline{\rho}({\bf r}_\alpha - \Delta l_i
\, \hat{\bf u}_i)
\Big) \, .
\eeq
The derivatives of the the single-particle energy~(\ref{epnu})
yield
an~expression for the velocity, as an analogous average to that
for the energy,
\beq
{\bf v} ({\bf p} ,{\bf r} ) = \sum_\alpha S_\alpha({\bf
r} - {\bf r}_\alpha) \, \overline{{\bf v}}( {\bf p} , {\bf
r}_\alpha )  \, ,
\eeq
and an expression for the force, amounting to a~prescription
for the gradient,
\beq
-{\partial \epsilon ({\bf p}, {\bf r})  \over \partial {\bf r}}
= - \sum_\alpha {d \, S_\alpha({\bf r} - {\bf r}_\alpha) \over
d{\bf r}} \,  \overline{\epsilon}( {\bf p} , {\bf
r}_\alpha )  \, .
\eeq

In the simulations, we take $S({\bf r}) = \prod_i g_i
(r_i/\Delta l_i)$.  In the interior of our computational area,
we
use $g(x) = 0.5$ for $|x| < 0.5$, $ g(x) = 0.75 - 0.5|x|$ for
$0.5 < |x| < 1.5$, and $g(x) = 0$ for $1.5 <  |x|$.
At a~forward edge, we use $g(x)=0.75 + 0.5 x$ for $-1.5 < x <
0.5$, and $g(x)=0$ outside of that interval.
The~accuracy of energy conservation for momentum-dependent
fields, following the
lattice Hamiltonian method, is illustrated in
Table~\ref{enco}.

\subsection{Collision Rates}

Different expectations might be held with regard to in-medium
scatterings.  For~perturbative processes, matrix elements in
the cross section would
be expected to be similar in medium to those in free-space.
However, for
strong interactions, with a~number of partial waves
participating, the~cross section might be invariant as
a~geometric quantity, if not for the fact that
the~space for the independent scatterings might get limited, in
a~dense medium.
We explore two types of cross sections in the medium:
the free cross sections and the cross-sections reduced
in such a manner that their radii are limited by
the interparticle distance,
\beq
\sigma = \sigma_0 \, \mbox{tanh} (\sigma_{free}/\sigma_0) \, ,
\eeq
where $\sigma_0 = \rho^{-2/3}$.  The~reduction is only applied
to the elastic cross sections, in order to maintain the
detailed
balance relations in the medium.  Some support for the reduced
cross sections stems from measurements of the linear momentum
transfer and of the ERAT cross sections~\cite{dan99}.
It may be mentioned, regarding the self-consistent
nuclear-matter calculations, that
a~reliance on the quasiparticle
approximation excludes, in~practice, the ability to see any
changes in
the scattering rates due to the overlap of scattering regions;
it is necessary to incorporate
a~spreading of the states~\cite{dic98,boz99}.

The momentum dependence of the MFs can make the
calculation of the collision rates difficult because of the
deformation of the energy shell.  We adopt a~simplifying
approximation of $d \epsilon / dT \approx$~const,
where~$\epsilon$ and~$T$ (kinetic energy) pertain to the same
momentum,
for collisions that locally contribute mostly to the transport.
Our approximation is a~relativistic analog of the effective
mass approximation.  Its validity, throughout the energy
range, is suggested by the
slowing of the variation in the MF
and in the optical potential as the energy increases, cf.\
Eq.~(\ref{Uopt}) and Fig.~\ref{upo1}.
Within the approximation, we can calculate the rates following
the
same procedure as for momentum-independent fields~\cite{dan91}
(see also~\cite{lan93}), colliding test-particles within the
cells between the
nodes, renormalizing only the rates for the average change in
velocities.
One way of assessing the quality of the approximation is by
examining the accuracy of energy conservation.  It is seen
in Table~\ref{enco} that the error does not exceed that from
integrating the MF motion; in some cases the two errors
appear to compensate.

\section{Elliptic Flow}

\subsection{Flow and the Momentum Dependence of the Mean Field}

At few hundred MeV/nucleon,
the best probes for the average forces in heavy ion
colisions are at present the
{\em directional features} of the transverse collective motion.
The~time for the development of the directional features is
limited by the passage of spectators near the reaction zone.
The~overall strength
of the motion is not such a~good probe
because the time for the development of the motion as a~whole
is not comparably limited.  Weaker forces in the matter are
likely to
be correlated with longer expansion times and stronger forces
with
shorter times, with the difference in the times compensating
for
the difference in the forces in the net strength of
motion.
Besides the size of the acting forces, an~issue in the
reactions
is the nature of these forces.  The~forces may be
associated with~MF, but the matter can also be pushed by
thermal pressure.  Varying contributions were found
in early central-reaction simulations~\cite{gal90} utlilizing
different
combinations of density and momentum dependence of the~MF.
A~weak momentum dependence allows for a~greater
increase in density and a~better equilibration and, hence,
for a~greater role for the thermal pressure.  In~assessing the
momentum dependence, we shall try to circumvent the question of
equilibration in central collisions and to follow up on that
question in
the light of the new results, as well as to reexamine EOS, in
the future.

The~flow anisotropies may be quantified in terms of the average
transverse
momentum component in the reaction plane or in terms of
the eigenvalues of the transverse momentum tensor, as functions
of
rapidity.  More flexibility in the quantification give
the moments of the azimuthal angle relative to the reaction
plane,
\beq
v_n = \langle \cos{(n \phi)} \rangle \, ,
\eeq
as these can be
investigated both as functions of rapidity and transverse
momentum.  At~midrapidity, away from spectator contributions,
the~lowest nonvanishing moment in symmetric systems is~$v_2$
describing ellipticity of the particle distribution;
the derivative of $v_1$, $d  v_1 /dy$, may be studied in that
region, too.

Figure~\ref{v2b} shows the dependence of
the midrapidity $v_2$
on impact parameter, obtained in the Boltzmann-equation
simulations of Au + Au collisions at 400~MeV/nucleon.
Mean fields with and without momentum dependence, corresponding
to different
incompressibilities, were utilized in those simulations.
The~negative values
of~$v_2$ for all simulations indicate a~preference for particle
emission out
of the reaction plane, towards 90$^\circ$ and~270$^\circ$.
For low impact parameters, the system approaches
azimuthal symmetry in space and the magnitude of $v_2$
decreases.
In~semicentral collisions, $b \lesssim 6$~fm, the~values
of~$v_2$ from simulations with different~MFs
cluster differently at different impact parameters, reflecting
an~interplay of the geometry, of the~dependence of~MF on
density and momentum, and of the thermal
pressure.  Beyond that lower-impact parameter region, however,
the~results of simulations
utilizing MF with and without momentum dependence, exhibiting
distinctly dissimilar behavior, clearly
separate.
Without momentum
dependence in the~MF, the values of $v_2$ first stabilize and
then
drop in magnitude.  With the momentum dependence, the~values of
$v_2$ continue to increase in magnitude up to very high impact
parameters, in a~roughly linear fashion.  These results give
rise
to an~expectation that the measurements of ellipticity
at high impact parameters
can be used to test the momentum
dependence of the~MF in collisions.

Due to its very nature, one might further hope that the
momentum dependence of MF would be revealed in the features of
emission of particles with high momenta.  In~the past, in
fact, we have indicated~\cite{pan93} that the behavior of the
first-order flow with transverse momentum, $v_1 (p^\perp)$,
could be used to discern the momentum-dependent from
the momentum-independent MFs in collisions.  To our knowledge,
though, no experimental analysis of those coefficients has been
done within the energy range where the degrees of freedom,
playing a~role in the collisions, are still under some control.
However, the~intermediate-energy data on the dependence of
ellipticity on~$p^\perp$
exist~\cite{lam94,bri96} and, following the discussion, one
might  hope to access
the momentum dependence of the~MF in $v_2$ at
high~$p^\perp$ and~$b$.

\subsection{Data Comparisons}

Figure~\ref{rnkt} compares the $p^\perp$-dependence of the
measured and calculated
ratios of out-of-plane to in-plane midrapidity proton-yields
($0.35 < y/y_{beam} < 0.65$), $R_N = (N(90^\circ)
+ N(-90^\circ))/(N(0^\circ)+N(180^\circ))$, in
400~MeV/nucleon $^{209}$Bi +
$^{209}$Bi collisions at~$b \simeq
8.7$~fm, as~computed from the lowest Fourier coefficients,
$R_N
= (1 - v_2)/(1 + v_2)$.
We utilize the KaoS~\cite{bri96} data in our comparison, rather
than the nominally similar LAND~\cite{lam94} data,
because of the wider available impact-parameter range
and because of an~absolute normalization of the KaoS data.
The~top panel of Fig.~\ref{rnkt} compares the data (squares) to
the calculations
(lines) done for in-medium cross sections and different MFs
which all correspond to the incompressibility of $K=210$~MeV.
The~experimental ratio rises rapidly with the transverse
momentum above 300~MeV/c, reaching values higher than 2
above 500~MeV/c.
The~ratio from simulations without momentum dependence in the
MF stays relatively flat with values below~1.4 up to momenta
of~700~MeV/c.  It~may be mentioned that the continuity of
the momentum distribution enforces~\cite{dan95}
a~quadratic behavior of~$v_2$, and thus of~$R_N$,
around~$p^\perp = 0$.
With the momentum dependence in the~MF, a~strong sensitivity
to the details in that dependence is observed
in the $p^\perp$-dependence of~$R_N$, with
the KaoS data at the displayed momenta favoring
the~MF parametrization characterized by
$m^*/m = 0.70$.  The~validity of conclusions on the
momentum dependence hinges on the sensitivity of the results to
the incompressibility and to cross sections.  The~sensitivities
are tested in the bottom panel of Fig.~\ref{rnkt}.
Simulations with MFs without momentum dependence, corresponding
to different incompressibilities, and the cascade model yield
all practically the same results for $R_N(p^\perp)$ at
high~$p^\perp$.  A~sensitivity to the in-medium NN cross
sections is observed, but it is weak in the high-momentum
region.

We next examine whether similar conclusions can be drawn at
other beam energies.  Thus, the~top panel of Fig.~\ref{rnk7t}
compares the $p^\perp$-dependence of the
measured and calculated midrapidity
ratios
in
700~MeV/nucleon $^{209}$Bi +
$^{209}$Bi collisions at~$b \simeq
8.6$~fm.  Again, the~ratios from simulations with
momentum-independent MFs are quite incompatible with the data.
As~to the details in the momentum dependence, the~region of
high sensitivity to those details shifts towards
higher transverse momenta, compared to collisions
at~400~MeV/nucleon.  We~should indicate that our ability to
address very high transverse momenta is
limited by the finite statistics in the simulations.
The~data, nonetheless, eliminate such weak momentum dependence
as that represented by the $m^*/m= 0.79$ parametrization.
Furthermore, the~700~MeV/nucleon data marginally favor
the parametrization characterized by $m^*/m=0.65$ over
$m^*/m=0.70$.
It~is seen in Fig.~\ref{rnk7t}, that some sensitivity to just
the density dependence of the MF
develops at~700~MeV/nucleon in the midperipheral collisions.

Some further shift of the region in $p^\perp$ where ellipticity
strongly discriminates different momentum-dependences of the~MF
is observed in the 1000~MeV/nucleon Bi + Bi collisions, see the
top panel of~Fig.~\ref{rnk10t}.  While the predictions
for the momentum-dependent~MFs clearly separate from those for
the
momentum-independent~MFs and are, by far, more consistent with
the data~\cite{bri96}, the~separation between the predictions
for the
$m^*/m \le 0.70$ and $m^*/m = 0.79$ MFs is not that
large at $p^\perp < 1000$~MeV/c.

Each time, so far, we compared the theory to data at about the
same impact parameter at the three beam energies.  The~bottom
panel of Fig.~\ref{rnk10t} illustrates the quality of data
description, by the calculation with~$m^*/m = 0.70$, at~different
impact parameters.  We should mention that we refrain from
deciding on the momentum dependence of the MF based on
comparisons to the highest available impact
parameters~\cite{bri96}, as nonlinearities in the variation of
quantities under the averaging over
impact parameter in the marginal region may affect the
results.
We should also mention that for some combinations of impact
parameter and beam energy, the values of
out-of- to in-plane ratios $R_N$ from the KaoS
measurements~\cite{bri96} fall below~1 at low~$p^\perp$.  We
do not observe such a~behavior in the calculations.
The~preliminary FOPI data on the $p^\perp$-dependence of
elliptic flow do not exhibit such a~behavior,
either~\cite{and99}.

Summarizing the analysis of simulations and the comparisons to
data thus far, we find that the ellipticity at high impact
parameters and transverse momenta is a~very sensitive probe
of the momentum dependence of the nucleonic~MF.  The large
measured values of out-of-plane to in-plane ratios
(or~ellipticity) at
midrapidity cannot be explained without the
momentum-dependence.
Selection of data favors an~MF parametrization
characterized by $m^*/m \simeq 0.70$.
Coincidentally, such parametrization agrees with the
information on the nucleonic~MF at~$\rho_0$ from nucleon
scattering.  A~natural question to ask at this stage is whether
the measurements of the ellipticity represent just another way
of
accessing the same information or whether these measurements
have
the potential to expand on the information from
nucleon scattering.

A~surmise regarding the same information as from scattering
might
be based on the fact that the densities reached in peripheral
collisions are lower than in the central~\cite{dan95}.
In~simulations with the momentum-dependent~MF at $b \sim
8.7$~fm, we find, though, the~maximal densities of
$\rho_{max}/\rho_0 \approx 1.85$, 2.20, and~2.40, at~400, 700,
and~1000~MeV/nucleon,
respectively.  However, the~spatial volume with such densities
might not be large or time span for the densities long enough
to affect the particle emission, or otherwise the
maximal densities might not be relevant.  To~assess whether the
elliptic flow tests the momentum dependence more at supranormal
or at normal and subnormal densities, we have carried out the
simulations with an~$m^*/m=0.70$ and $K=210$~MeV MF modified
by making the momentum dependence at $\rho > \rho_0$ follow the
dependence at~$\rho_0$.  The~modification was accomplished by
demanding that the velocity in the matter in (\ref{emo=})
ceased to change (i.e.\ increase) beyond the normal density:
\beq
v_X^*(p,\xi) = v_X^*(p,1) \, , \hspace*{2em} \mbox{for
$\xi=\rho/\rho_0  > 1$} \, .
\eeq
As the freezing of the momentum dependence softens the
zero-temperature energy density above~$\rho_0$,
we compensated that softening by
adding a~repulsive term to the potential~$U$ in~(\ref{emo=})
and~(\ref{U=}), of the form 25~MeV$\, (\xi-1)^{0.35}$, for
$\xi > 1$, roughly
restoring the density dependence of the energy from before the
modification.  Note that freezing the momentum dependence at
zero rather than normal density would yield
a~momentum-independent~MF.
The~results of the simulations with the
momentum dependence of MF frozen above~$\rho_0$ are shown in
the bottom panel of~Fig.~\ref{rnk7t}.  It~is seen that the
yield ratios $R_N$ are now well below data at high transverse
momenta.
The~results are, in~fact, a~bit closer to the results obtained
with the momentum-independent~MF than to the results obtained
with the MF with the standard
dependence.  Clearly, the~elliptic flow tests the
momentum dependence at supranormal densities.

In~the view of the last finding, the~preceeding comparisons to
data at 700 and 1000~MeV/nucleon and at high transverse momenta
suggest that the momentum dependence in the so-far favored
$m^*/m
=0.70$ MF parametrization may not actually grow fast enough
with the density at the highest reached densities and momenta.
We next investigate reasons for the sensitivity to MF at high
densities.

\subsection{Dynamics and Anisotropy at High Transverse Momenta}

\begin{figure}
\centerline{\includegraphics[angle=180,
width=.45\linewidth]{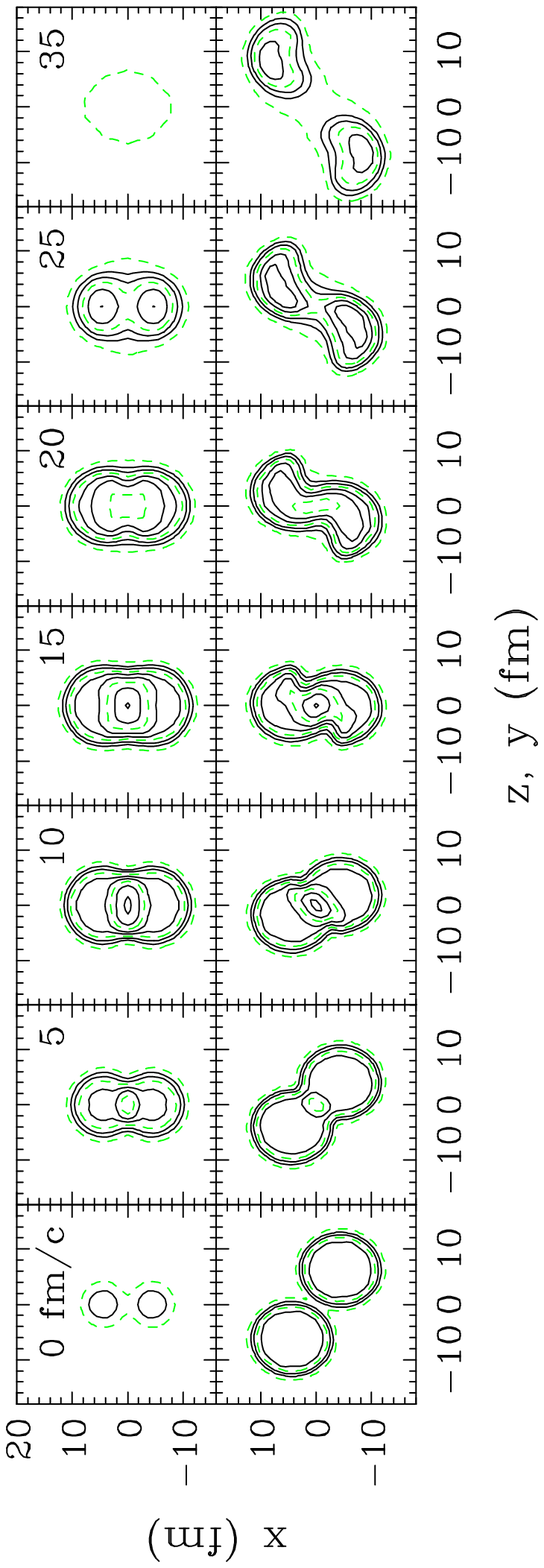}}
\vspace*{.1in}
\caption{
Contour plots of baryon density in the 400 MeV/nucleon Bi + Bi
reaction at $b=8.7$~fm in the reaction plane (bottom panels)
and in the plane through the system center othogonal to the
beam axis (top panels).  The~simulation was done using the MF
characterized by
$K=210$~MeV and $m^*/m = 0.70$.  The~density contours are shown
at intervals
of 0.2 in $\rho/\rho_0$, starting from $\rho/\rho_0 = 0.1$.
Every third contour, i.e.\ for $\rho/\rho_0 =0.1$, 0.7,
and~1.3, is represented by a~dashed line.  The~remaining
contours are represented by solid lines.
Numbers in the figure indicate time in~fm/c.
}
\label{conbb}
\end{figure}

To understand the association of high-momentum
anisotropies with the momentum dependence of MF at high
densities, we examine the production
rate for high-momentum
midrapidity protons in midperipheral collisions.  In~the upper
panel of Fig.~\ref{burst},
we plot the emission rate as a~function of time for protons
with
$p^\perp > 550$~MeV/c in the $K=210$~MeV $m^*/m=0.70$ Bi + Bi
simulations at 400~MeV/nucleon and $b=8.7$~fm; we also plot in
that figure the density at the system center.
The~emission time is defined as the instant of last
collision.
It is seen that the peak emission is strongly correlated with
the
system reaching a~high density at the center.  For comparison,
we
plot in the bottom panel of~Fig.~\ref{burst} the emission rate
for all midrapidity protons.  The~emission for all protons only
sets in during the high-density stage.  Thus, at~$t=30$~fm/c,
when 90\% of all high-momentum protons have already been
emitted and central density has fallen to low values, fewer
than 30\% of all midrapidity protons have been emitted.
Of all midrapidity protons, these are the high-$p^\perp$
protons that are most directly emitted from the high-density
zone during the overlap of the nuclei.  For a~view of
the overall progress of the Bi + Bi reaction,
see~Fig.~\ref{conbb}.

A~possible cause for the large high-$p^\perp$ anisotropies
obtained with the momentum-dependent MFs, larger than with the
momentum-independent MFs, might be a~stronger correlation of
the high-$p^\perp$ emission with the overlap of the nuclei for
the momentum-dependent
fields.  The~spectator pieces are, generally, expected to
shadow the emission within the reaction plane.
To~clarify the issue, in addition to the high-$p^\perp$ rate
for the momentum-dependent MF, we plot in the top panel of
Fig.~\ref{burst} the high-$p^\perp$ emission rate obtained with
the $K=380$~MeV
momentum-independent~MF.  While there are some statistically
significant differences between the rates, their overall time
dependencies are fairly similar; in~no way the small
differences could explain the huge differences in anisotropies
at high~$p^\perp$ found in~Fig.~\ref{rnkt}.  The~time
dependencies for the central densities and for the density
distributions
are quite similar for the two MFs, difficult to distinguish by
eye.  Thus, some differences in the directionality of the
emission
for the two MFs must exist at the same emission times, and be
responsible for the large ratios
in~Figs.~\ref{rnkt}-\ref{rnk10t}.

To more understand the emission, we shall examine $v_2 =
\langle \cos{2\phi} \rangle$, for the high-momentum protons, as
a~function of time.  Before that, we need to discuss
expectations regarding~$v_2$ in simple situations.  For
emission strictly towards 90$^\circ$ and~270$^\circ$ relative
to  the
reaction plane, we expect~$v_2=-1$.  For a~model
emission
from rough flat surfaces with normal directions towards
90$^\circ$ and 270$^\circ$, as suggested by the 10--20~fm/c top
panels
in~Fig.~\ref{conbb},
the~emission pattern is $dN/d\phi \propto |\sin{\phi}|$.
The~value for the coefficient is then $v_2=-1/3$.
(The~roughness scale, that should be much smaller than the
surface size, is the mean-free-path in our case.)
Another
interesting value of~$v_2$ is that for the most anisotropic
distribution describable in terms of two Fourier cofficients,
i.e.~$dN/d\phi \propto (1-\cos{2\phi})$, for which~$v_2 =
-1/2$.

\begin{figure}
\centerline{\includegraphics[angle=0,
width=.70\linewidth]{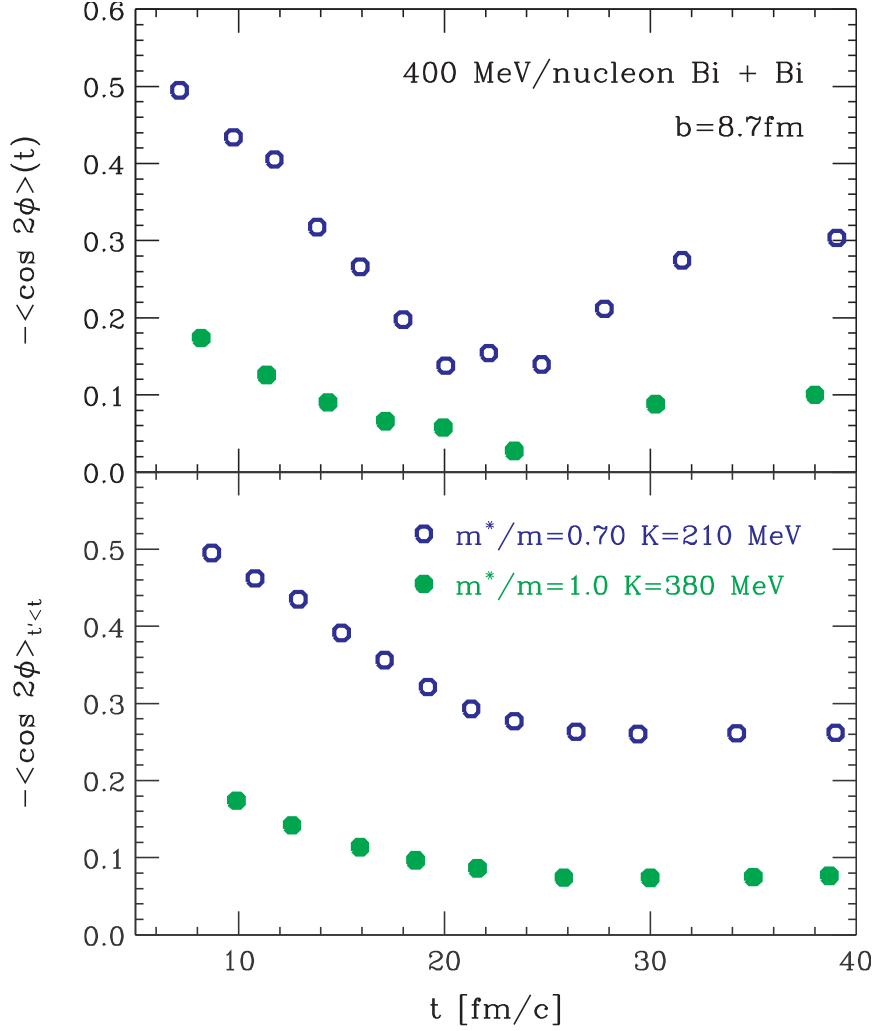}}
\vspace*{.1in}
\caption{
Time dependence of $-v_2$ for protons at $p^\perp >
550$~MeV/c from 400~MeV/nucleon Bi + Bi simulations employing
either the $m^*/m = 0.70$ $K=210$~MeV MF (open circles) or
the $K=380$~MeV momentum-independent~MF (filled circles).
The~top panel shows the average values for protons emitted in
the vicinity of a~given time, while the bottom panel shows the
value for all protons emitted up to that time.
}
\label{co2}
\end{figure}

Figure~\ref{co2} shows the time-dependence of $-v_2$ for
high-$p^\perp$ protons from the 400~MeV/nucleon Bi + Bi
simulations at $b=8.7$~fm employing the two recently
discussed~MFs.  Both the average values (top panel) for protons
emitted in the vicinity of a~given time, $-\langle \cos{2 \phi}
\rangle(t)$, as well as the values
(bottom panel) for all protons emitted up
to a~given time, $-\langle \cos{2 \phi} \rangle_{t'<t}$, are
shown.
It is apparent from Fig.~\ref{burst} and from the bottom panel
of~Fig.~\ref{co2} that the values of~$v_2$ for particles
emitted past $\sim 25$~fm/c have already little effect on the
overall value for all particles.

As is seen in the top panel of Fig.~\ref{co2}, the~emission for
the momentum-independent MF starts out with values of $-v_2
\sim 0.15$, in
the vicinity of a~half of what is expected for a~flat surface
parallel to the reaction plane.  It appears that the~emission
geometry is more complicated than in the latter case, with
a~finite size of roughness for the surface and with the surface
oriented towards different angles as may be apparent from the
bottom  5--15~fm/c panels in Fig.~\ref{conbb}.
In contrast to the momentum-independent~MF,
the~emission for the
momentum-dependent~MF
starts out with much higher values of $-v_2 \sim
0.50$,  well above the expectation for a~flat surface.  As
geometry
is virtually the same as for the first~MF, it must be
a~difference in the field within which the particles move
that plays a~focussing role.  Basically, the~effect may be
understood in terms of a~relatively strong repulsive~MF felt by
the high-momentum particles in the second case in the
high-density zone.  As
the particles move out, they feel a~strong gradient towards the
normal of the~emitting surface.  This gradient pushes out the
otherwise rapidly falling distribution of emitted particles in
the transverse
direction.  The~highest transverse momenta are acquired by
particles best aligned with the gradient which gives rise to
the enhanced anisotropy at the high momenta.
As the spectator pieces pass by the center of the system,
there is a~burst of particles that
were moving in the transverse direction parallel to the
reaction in the participant zone and were trapped till then.
That lowers the instantaneous $-v_2$ values in Fig.~\ref{co2}
around $t \simeq 20$~fm/c for either of the employed MFs.
Later, the instantaneous values recover somewhat, possibly
due to left-over particles from the early stage that just
underwent some secondary collisions and got reclassified with
regard to their emission time.

\begin{figure}
\centerline{\includegraphics[angle=90,
width=.95\linewidth]{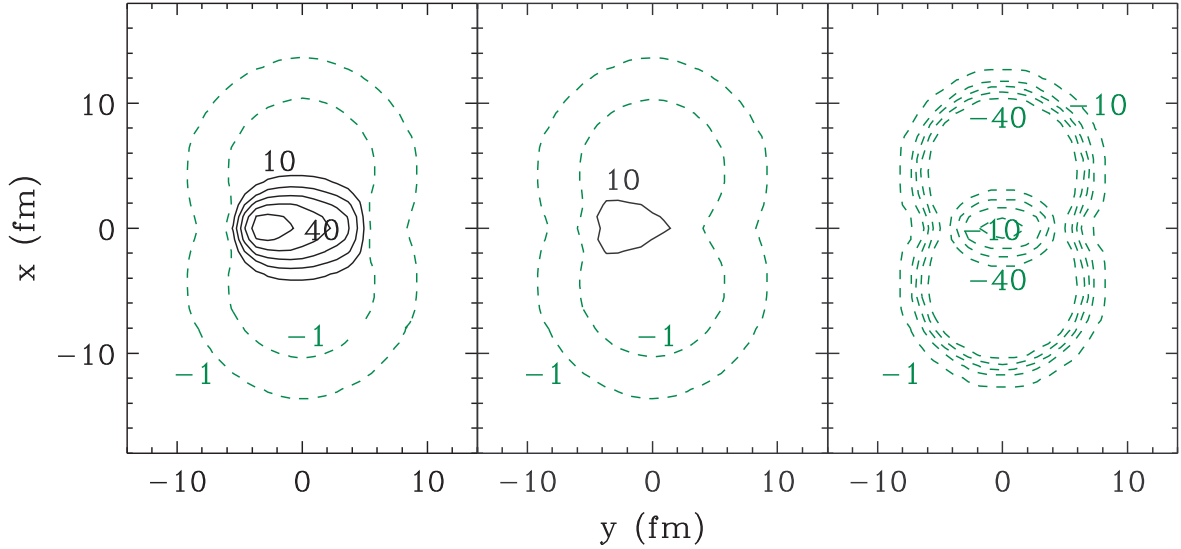}}
\vspace*{.1in}
\caption{
Contour plots of the optical potential field,
Eq.~(\protect\ref{Uopt}),
in the 400 MeV/nucleon $b=8.7$~fm Bi + Bi
reaction,
within the plane through the system
center perpendicular
to the beam axis, at $t=10$~fm/c, for nucleons with
$p^\perp = 600$~MeV/c directed to the right of the plot.
The~three panels,
from left to right of the plot respectively, show results for
three different~MFs: the standard MF characterized by
$m^*/m=0.70$ and $K=210$~MeV, the~same MF with
momentum-dependence frozen above~$\rho_0$, and the
momentum-independent MF characterized by $K=310$~MeV.
The~Coulomb and isospin contributions to the potential are
ignored in every case.  The~contours are indicated for the
potential value  of $-1$~MeV and for positive and negative
multiples of~10~MeV.  The~contours for the positive and
negative values are represented, respectively, by the solid and
dashed lines.  The~$-1$, $\pm 10$, and $\pm
40$~MeV contours are labeled.
}
\label{conu}
\end{figure}

\begin{figure}
\centerline{\includegraphics[angle=0,
width=.80\linewidth]{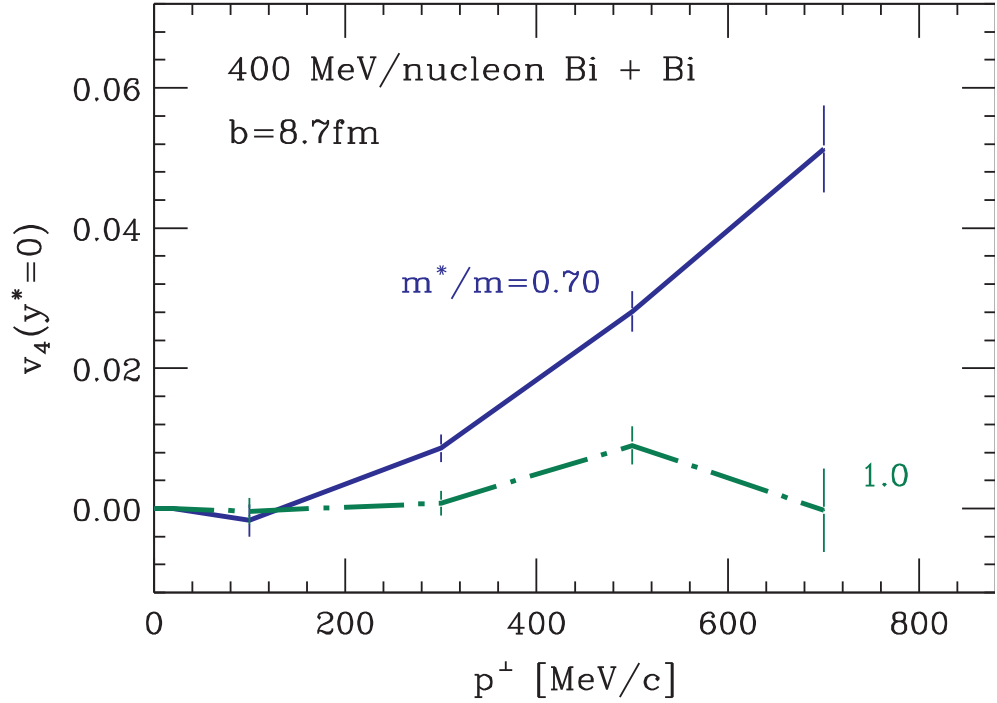}}
\vspace*{.1in}
\caption{
Proton Fourier coefficient $v_4 = \langle \cos{4 \phi}
\rangle$ as a~function of transverse momentum at midrapidity in
the Bi + Bi system at 400~MeV/nucleon and $b=8.7$~fm, from
simulations utilizing the $m^*/m=0.70$ $K=210$~MeV MF (solid
line) and the $K=380$~MeV momentum-independent MF (dash-dotted
line).
Numbers in the figure indicate the effective mass.
}
\label{v4t}
\end{figure}

Optical potential fields at a~high-density stage, in
simulations of the Bi + Bi reaction employing different MFs
are shown in Fig.~\ref{conu} for high-$p^\perp$ nucleons.
The MFs include the momentum-independent MF characterized by
$K=380$~MeV, the~MF characterized by $m^*/m = 0.70$ and
$K=210$~MeV
MF, and that last MF with the momentum dependence frozen
above~$\rho_0$.  The~MF with frozen momentum dependence
fails to describe the magnitude of elliptic anisotropy at
high~$p^\perp$ in the semiperipheral Bi + Bi at 700~MeV/nucleon
and, likewise, at~400~MeV/nucleon.
In~the high-density zone,
the~optical potential for $p^\perp = 600$~MeV/c nucleons is
strongly
repulsive for the standard~MF, reaching values of $\sim
55$~MeV.  Still higher
values are reached for higher momenta.  Notably,
in normal matter, the~optical potential for that MF, as well as
the potential from nucleon scattering, vanishes in the vicinity
of $600$~MeV/c (see also Table~\ref{tabmo}).
In the case of the frozen momentum
dependence, the~optical potential in the high-density zone
reaches only~$\sim 15$~MeV.
Finally, for the momentum-independent~MF, the~optical potential
stays mildly attractive at 400~MeV/nucleon in the high-density
zone, reaching maximal values there in the vicinity of~$\sim
-5$~MeV.  The~attractive potential could principally play
a~defocussing role in the emission; the~net results for
anisotropy are, nonetheless, similar to the results
without~MF, as was demonstrated earlier.
The~buldging out of the momentum distribution, due to
a~repulsive field for high momenta, suggests the possibility
for positive values of the fourth Fourier-coefficient at
midrapidity, $v_4 = \langle \cos{4 \phi}\rangle$;
the~coefficient would account
for a~fine structure in the enhancement around 90$^\circ$
and~270$^\circ$.
The~coefficient is shown as a~function of transverse momentum
in Fig.~\ref{v4t} for the $m^*/m = 0.70$ $K=210$~MeV MF and
for the momentum-independent $K=380$~MeV~MF.  For the
momentum-dependent MF the coefficient rises indeed rather
rapidly with momentum, whereas it remains close to zero for the
momentum-independent~MF.

\begin{figure}
\centerline{\includegraphics[angle=0,
width=.75\linewidth]{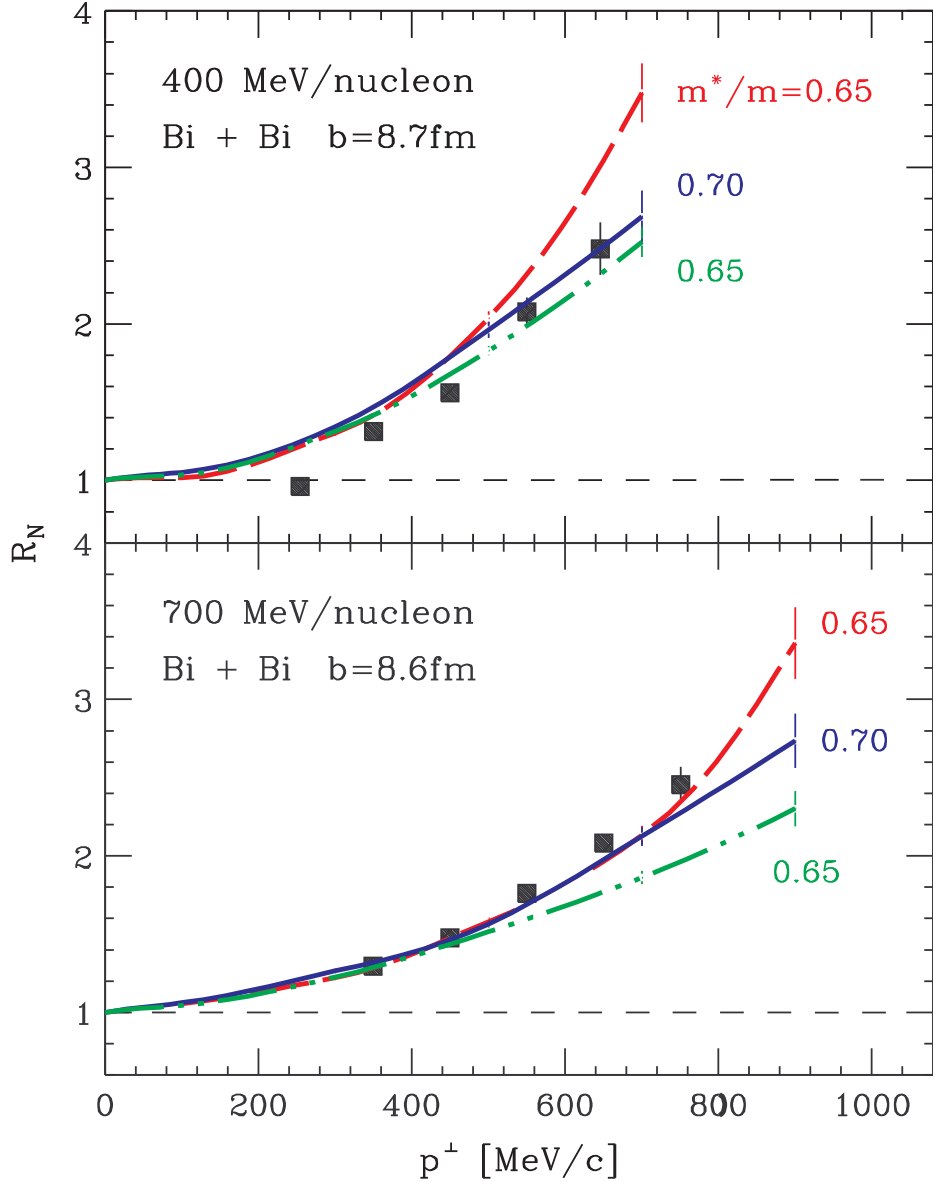}}
\vspace*{.1in}
\caption{
Measured~\protect\cite{bri96} (filled squares) and calculated
(lines) ratios
of out-of-plane to in-plane proton yields at midrapidity (0.35
$< y/y_{beam} <$ 0.65), as a~function
of transverse momentum, in
$^{209}$Bi +
$^{209}$Bi collisions at~$b =
8.6$~fm
and energies of 400 (top panel) and
700~MeV/nucleon (bottom panel).
Both the measured and calculated ratios are computed
from the
lowest Fourier coefficients of the proton distribution, $R_N
= (1 - v_2)/(1 + v_2)$.  The~calculations were carried out
using in-medium cross sections and MFs characterized by the
incompressibility of $K=210$~MeV.  The~numbers in the figure
indicate the values of the effective mass for the MFs;
the~solid lines represent
results obtained using the MF in Table~\protect\ref{tabmo}
characterized by $m^*/m = 0.70$ and the long-dashed and
long-dash-double-dotted lines represent results obtained the
first and second $m^*/m = 0.65$~MF in the table, respectively.
}
\label{rnk47t}
\end{figure}

One outstanding question, which we need to address, is whether
we can actually draw conclusions on the effective mass in
ground-state matter, be that in an~average sense.
The~effective mass was, so far, quite useful in labeling
the different~MFs.  To~address the question, we construct
an~MF parametrization requiring that the effective mass at
Fermi energy in the ground-state matter is $m^*/m = 0.65$ and
that the optical potential vanishes at the same momentum
as for our previous $m^*/m = 0.70$~MF, i.e.~at $p_0
\approx 680$~MeV/c, cf.~Table~\ref{tabmo}.  The~requirement
makes the
potentials appear similar at intermediate momenta; at
high momenta, though, the $m^*/m = 0.65$ potential becomes
softer than $m^*/m=0.70$, approaching a~lower asymptotic
value, see the table.  We now turn to the predictions for
ellipticity.
Figure~\ref{rnk47t} shows the calculated ratio of in to out-of
plane
proton emission at midrapidity as a~function of transverse
momentum in midperipheral Bi + Bi reactions at 400 and
700~MeV/nucleon, for the new and previous MFs, together with
the data.  At 400~MeV/nucleon, the~predictions obtained using
the
$m^*/m=0.70$~MF and the new $m^*/m=0.65$~MF are fairly similar.
With the increase in beam energy, however, the~predictions
separate, with these for the new $m^*/m=0.65$~MF falling
below those for the~$m^*/m=0.70$~MF and below the data.
Clearly,
at the considered energies, the~high-momentum
elliptic flow
is sensitive to the momentum-dependence of MF at intermediate
and high momenta but not at momenta as low as
the
ground-state Fermi momentum.  On the other hand,
given the requirements of the energy and density for the
ground state, and of the optical potential becoming
repulsive but not too repulsive with the increase in momentum
to yield right anisotropies, it is difficult to put forward
a~MF with $m^*$ outside of the interval $m^*/m \simeq
(0.63-0.73)$, unless that MF were to have rather discontinuous
behavior.  Of course,
on account of
the dispersion
relation involving single-particle energies and
scattering rates and the discontinuity
in the Pauli blocking at zero temperature,
there might be variations in $m^*$ in the
very vicinity of the Fermi momentum,
but these
would be quite local.  Now we turn to a~direct comparison of
our results to microscopic calculations.

\section{Comparisons to Microscopic Optical-Potential Calculations}

The best way to test the momentum-dependence in microscopic MFs
would be to implement
these MFs into a~transport calculation and confront
the elliptic-flow results with data.  Given though the
preliminary nature of
our investigation, where the elliptic flow is
exploited for the
first time in exploring the momentum dependence, we will merely
compare directly
the utilized optical potentials to those calculated in the
literature, at densities reached in reactions and at momenta
where anisotropies are observed.  That
will also give the
opportunity to take a~closer look at the potentials in
simulations, presented so far only for normal and subnormal
densities.
We shall comment on some difficulties in the comparisons on the
way.

\subsection{Optical Potentials from
Dirac-Brueckner-Hartree-Fock Calculations}

The first microscopic optical potentials that we consider are
those obtained within the
Dirac-Brueckner-Hartree-Fock (DBHF)
approach~\cite{har87,bro90}.  In that approach, an~in-medium
Thompson equation is solved with a realistic relativistic
interaction.  The~equation accounts for the effects of
Pauli principle and of self-consistent scalar and vector~MFs.
Rise in the scalar~MF with density leads to a~weakening of the
scalar exchange in the interaction and contributes to the
nuclear saturation~\cite{bro90}.
In Fig.~\ref{ugqli}, we compare the DBHF optical potentials
obtained using the Bonn-A interaction~\cite{bro90} to our
parametrizations.
The~potentials from a~relativistic approach
may be compared as functions of momentum.
The~short-dashed lines in the figure represent the potential
determined in~\cite{lim93,bro90} for a~broad range of densities
assuming momentum-independent scalar and vector~MFs.
The~short-dashed-dotted lines represent the optical potential
determined in~\cite{lee97} for a~more narrow range of densities
assuming a~parametrized form of the momentum dependence for the
scalar and vector~MFs. The~momentum-dependence of the optical
potential for
the momentum-independent MFs stems from a~large magnitude of
the two MFs that largely cancel out while differently entering
the
single-particle energies.  That effect also largely contributes
to the momentum dependence of the optical potential in the case
of momentum-dependent~MFs; see the similarity of the two
potentials in~Fig.~\ref{ugqli}.  The~solid and long-dashed
lines in the figure represent, respectively, our $m^*/m=0.70$
and first $m^*/m=0.65$ $K=210$~MeV parametrizations.

It may be seen in Fig.~\ref{ugqli} that, at densities and
high-momenta explored in the 400~MeV/nucleon
collisions, the~DBHF potential~\cite{lee97} is fairly close to
that from our $m^*/m=0.70$ parametrization.  The~DBHF
potential~\cite{lim93} would likely give a~too large
high-$p^\perp$ elliptic anisotropy at 400~MeV/nucleon.  As the
density
increases, the~DBHF potential~\cite{lim93} rises more and more
above our $m^*/m=0.70$ potential.  However, at the higher two
beam energies in Bi + Bi collisions, involving somewhat higher
densities, we did not
explore how strong the MF would need to be to yield excessive
anisotropies.  Thus, we cannot classify the behavior of the
DBHF potential at the increasing densities as unrealistic.
We should note that the optical potential in our $m^*/m=0.70$
parametrization appears more repulsive at $p=650$~MeV/c and
at $\rho/\rho_0 \sim 1.5$ in the simulation in Fig.~\ref{conu}
than
at such momentum at $\rho/\rho_0 = 1.5$ in Fig.~\ref{ugqli}.
Basically, the~effect may be understood in the following
way.\footnote{This type of an effect was discussed before, in
different terms, by Wolter \etal.~\protect\cite{fuc96}.}
As~particles
in a~reaction are locally at higher typical momenta
than in the~zero-temperature matter, they feel, on the average,
a~more repulsive potential.  By self-consistency this makes the
potential more repulsive at any fixed
momentum. We expect
such dynamic shifts for all the potentials.  In addition,
a~given fixed momentum in the overall frame corresponds to
different momenta in local frames in a~reaction which
is similar for different potentials in simulations.

One may notice in Fig.~\ref{ugqli} that, at
$\rho=\rho_0=0.16$~fm$^{-3}$ and momenta $p < 400$~MeV/c,
the~DBHF potentials are always lower than the potentials from
our parametrizations.
This might be due to
a~problem
in the DBHF calculations with the thermodynamic consistency
relation (Hugenholtz-van Hove theorem):
\beq
{e \over \rho} = \epsilon^F
\label{econ}
\eeq
in the ground state.  Our optical potentials cross at the Fermi
momentum in Fig.~\ref{ugqli}, as required by~(\ref{econ}).
The~fact is, however, that the DBHF potentials are a~bit too
attractive
at low momenta and the nuclear matter within the DBHF approach
saturates
at a~slightly excessive density of
0.18~fm$^{-3}$~\cite{bro90,lee97}.
Overall, though, the~proximity of the DBHF potentials,
especially for the momentum-dependent~MFs, to the
parametrization which adequately describes the elliptic-flow
anisotropies, is rather startling in the tested range of
densities and momenta.

\subsection{Potential by Baldo {\em et al.}}

\begin{figure}

\centerline{\includegraphics[angle=0,
width=.70\linewidth]{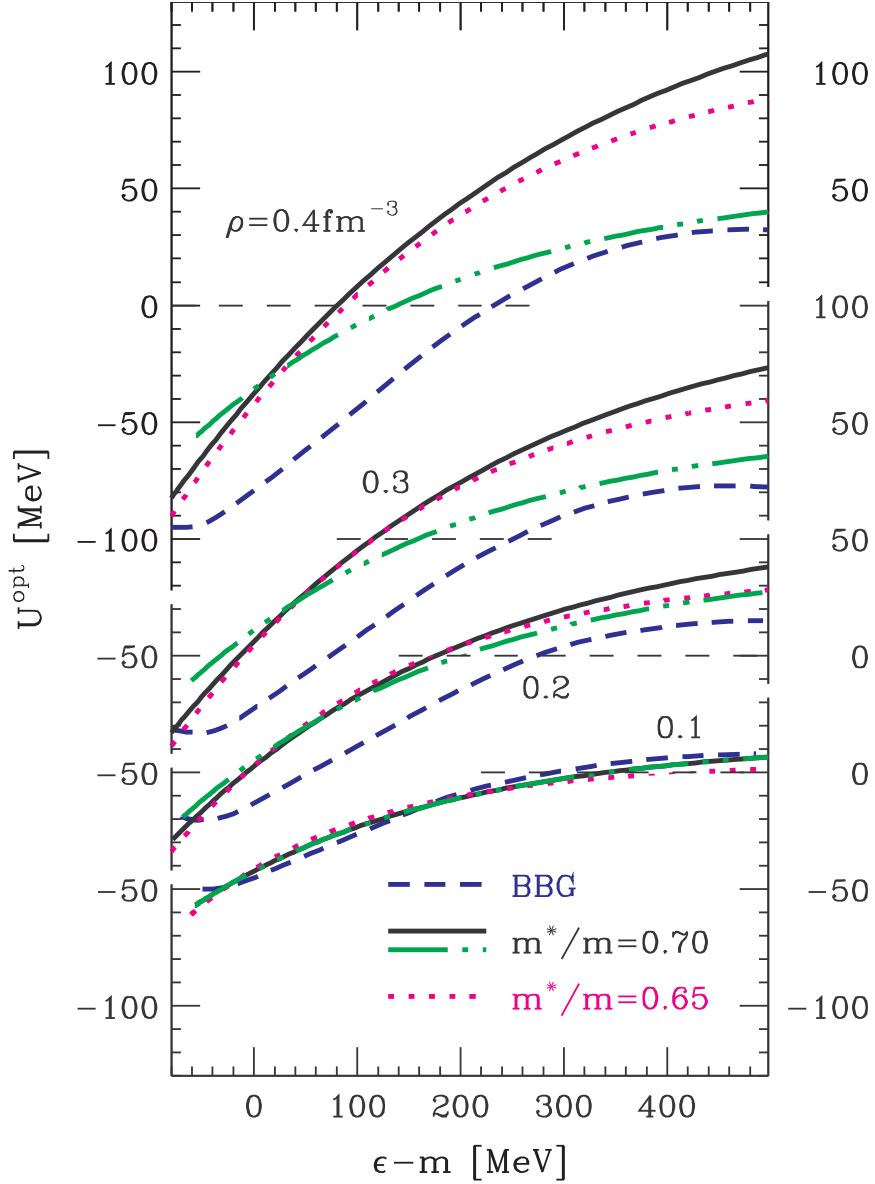}}
\vspace*{.1in}
\caption{
Optical potential in nuclear matter as
a~function of nucleon energy, at different densities, from
the BBG
calculations with Paris interaction~\protect\cite{bal89,ins94}
and in our parametrizations.
The~short-dashed, dotted, and solid lines represent the BBG
potential and the potential for our second $m^*/m=0.65$ and
$m^*/m = 0.70$ $K=210$~MeV parametrizations, respectively.
The~long-dash-double-dotted lines represent the potential for
our $m^*/m=0.70$
parametrization with the momentum dependence frozen
above~$\rho_0=0.16$~fm$^{-3}$.
The~numbers
in the figure indicate the values of density in units
of fm$^{-3}$.
The~thin horizontal dashed lines indicate the zero
value for the potential.
}
\label{ubal}
\end{figure}

Next potential, being considered, is that obtained within the
nonrelativistic Brueckner-Bethe-Goldstone (BBG)
approach with the Paris interaction~\cite{bal89,ins94}.
In~that approach, to~the Brueckner-Hartree-Fock term in the
optical potential
a~so-called rearrangement term is added which accounts for the
influence
of the particle,
for which the potential is evaluated,
on the correlations
between other particles in the medium.
At low momenta,
the~rearrangement term weakens the dependence of the optical
potential on momentum in quite an~essential manner.  The~BBG
potential, represented by a~short-dashed line, is compared to
our parametrizations as a~function of
energy at several densities in Fig.~\ref{ubal}.  It~is seen
that, at all the
supranormal densities in the figure and, essentially, at all
energies, the~BBG potential is below our second $m^*/m=0.65$
parametrization and even below the $m^*/m=0.70$ parametrization
with
the momentum dependence frozen at the higher densities;
the~potentials from the two parametrizations are represented by
the dotted and
long-dash-double-dotted lines, respectively.  Notably,
the~nuclear matter in
the BBG approach saturates at a~very high density of $\rho =
0.24$~fm$^{-3}$~\cite{ins94}.  From our investigations, we can
conclude that the BBG
potential is unacceptably weak at the densities $\rho \gtrsim
1.5 \rho_0$ and higher energies and momenta; that potential
would yield far too small elliptic anisotropies in the Bi + Bi
collisions at 400 and 700~MeV/nucleon.

\subsection{Potentials from Variational Calculations}

The final set of optical potentials, which we examine, is that
obtained within the variational method for nuclear matter
for three combinations of
two- and three-nucleon interactions~\cite{wir88}.
These combinations included the Friedman and
Pandharipande~\cite{fri81} combination of the Urbana $v_{14}$
(UV14)
with a~three-nucleon interaction (TNI), and the combinations of
either the Urbana
$v_{14}$ or the Argonne $v_{14}$ (AV14) nucleon-nucleon
interaction
with the Urbana model VII (UVII) three-nucleon interaction.
As criticized in~\cite{bal89}, the~rearrangement contribution
was not included in determining the optical potentials
in~\cite{wir88}; such a~criticism might be also raised
against the~DBHF calculations.
The~nuclear matter was constrained in the
calculations~\cite{wir88}
to saturate at $\rho=0.157$~fm$^{-3}$.  Earlier Friedman and
Pandharipande results (UV14+TNI) within the approach, for
subnormal
densities, were already displayed in Fig.~\ref{upoden}.

\begin{figure}

\centerline{\includegraphics[angle=0,
width=.70\linewidth]{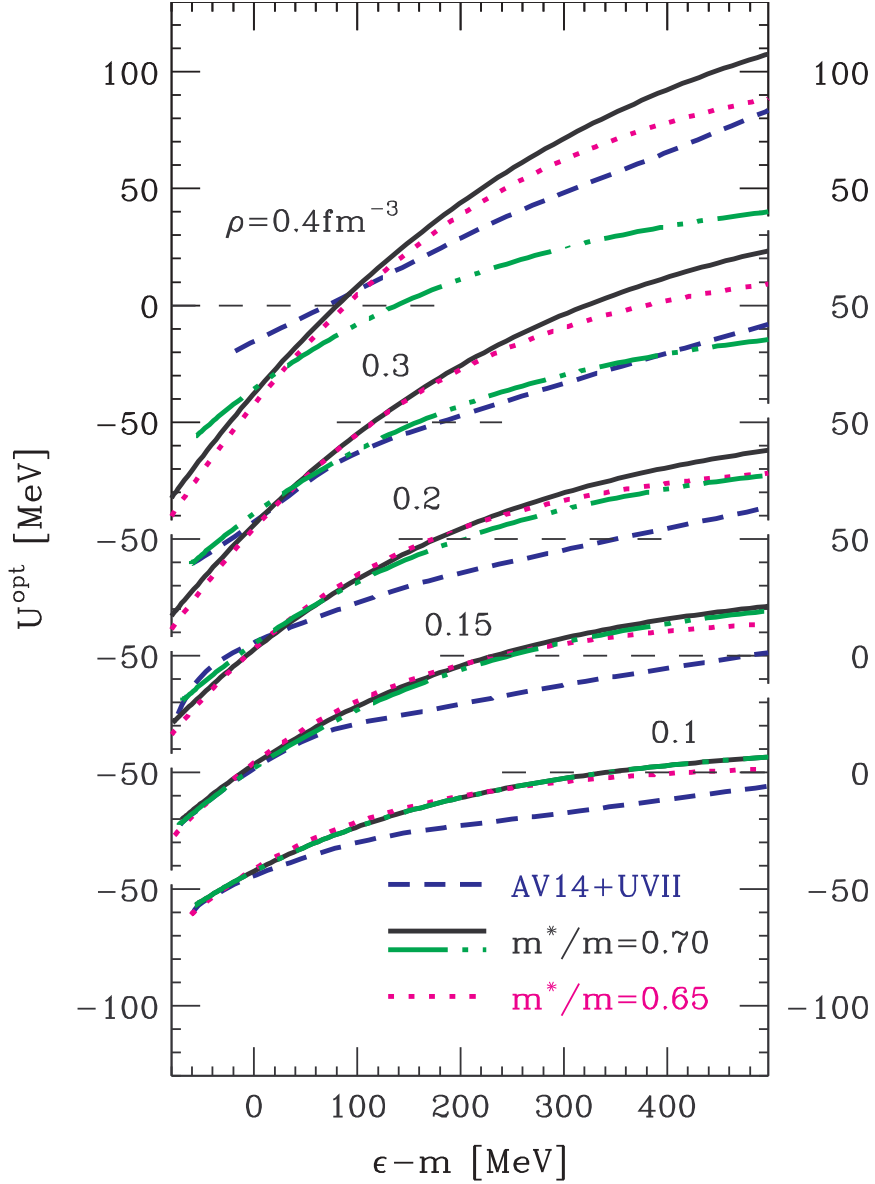}}
\vspace*{.1in}
\caption{
Optical potential in nuclear matter as
a~function of nucleon energy, at different densities, from
the variational
calculations with AV14 + UVII interactions~\protect\cite{wir88}
and in our parametrizations.
The~short-dashed, dotted, and solid lines represent the AV14 +
UVII
potential and the potential for our second $m^*/m=0.65$ and
$m^*/m = 0.70$ $K=210$~MeV parametrizations, respectively.
The~long-dash-double-dotted lines represent the potential for
our $m^*/m=0.70$
parametrization with the momentum dependence frozen
above~$\rho_0=0.16$~fm$^{-3}$.
The~numbers
in the figure indicate the values of density in units
of fm$^{-3}$.
The~thin horizontal dashed lines indicate the zero
value for the potential.
}
\label{uwir}
\end{figure}

Figure~\ref{uwir} compares, as a~function of energy and at
several densities, the~optical
potential~\cite{wir88} for the AV14 + UVII interactions
to the potentials for our $K=210$~MeV
MF parametrizations.
It is seen that, at the densities in the figure and at high
momenta, the~AV14 + UVII potential, represented by
a~short-dashed line, is always below the potential from our
second $m^*/m=0.65$ parametrization;
at $\rho \simeq 1.7
\rho_0$,
it is even below the $m^*/m=0.70$ parametrization with
the momentum-dependence frozen.
This is then similar
to the case of the BBG potential.  The~AV14 + UVII potential is
unacceptably weak at supranormal densities.  Use of this
potential in simulations would lead to a~serious
underestimation of high-momentum elliptic anisotropies in the
semiperipheral Bi + Bi collisions at 400 and at
700~MeV/nucleon.

Finally, we turn to UV14 potentials.  Both the UV14 + TNI
and UV14 + UVII potentials~\cite{wir88} are compared to the
potentials from
our parametrizations in Fig.~\ref{uwirb}.  It is seen that at
supranormal densities in the figure, the~UV14 + TNI potential,
represented by a~short-dashed line, is very close to the
potential from our $m^*/m = 0.70$ parametrization, represented
by a~solid line.  The~UV14 + UVII potential, represented by
a~solid-dash-dotted line, is below the potential from our
second $m^*/m=0.65$ parametrization, represented by a~dotted
line, at $\rho \lesssim 2 \rho_0$.
Thus, a~use of the UV14 + UVII in a~simulation should lead to
some underestimation of the high-$p^\perp$ elliptic anisotropy
in the Bi + Bi collision at 700~MeV/nucleon.

\section{Discussion}

\begin{figure}
\centerline{\includegraphics[angle=0,
width=.75\linewidth]{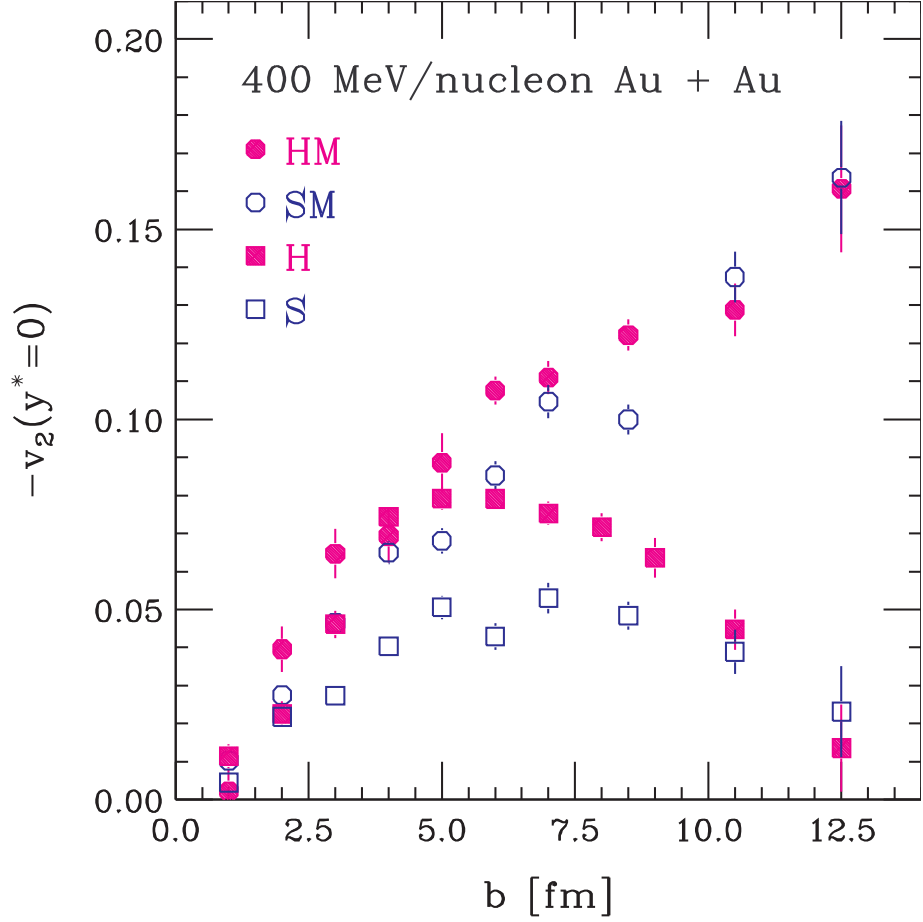}}
\vspace*{.1in}
\caption{
Negative of the ellipticity coefficient
at midrapidity,
as a~function of the impact parameter,
from simulations of Au + Au collisions at 400~MeV/nucleon.
The~squares and circles represent, respectively, the~results
for MF without and with momentum dependence
($m^*/m = 0.65$, of~Ref.~\protect\cite{dan98}).  The~filled
symbols
are for the incompressibility~$K=380$~MeV and the open symbols
are for~$K=210$~MeV.
}
\label{v2b}
\end{figure}

\begin{figure}
\centerline{\includegraphics[angle=0,
width=.75\linewidth]{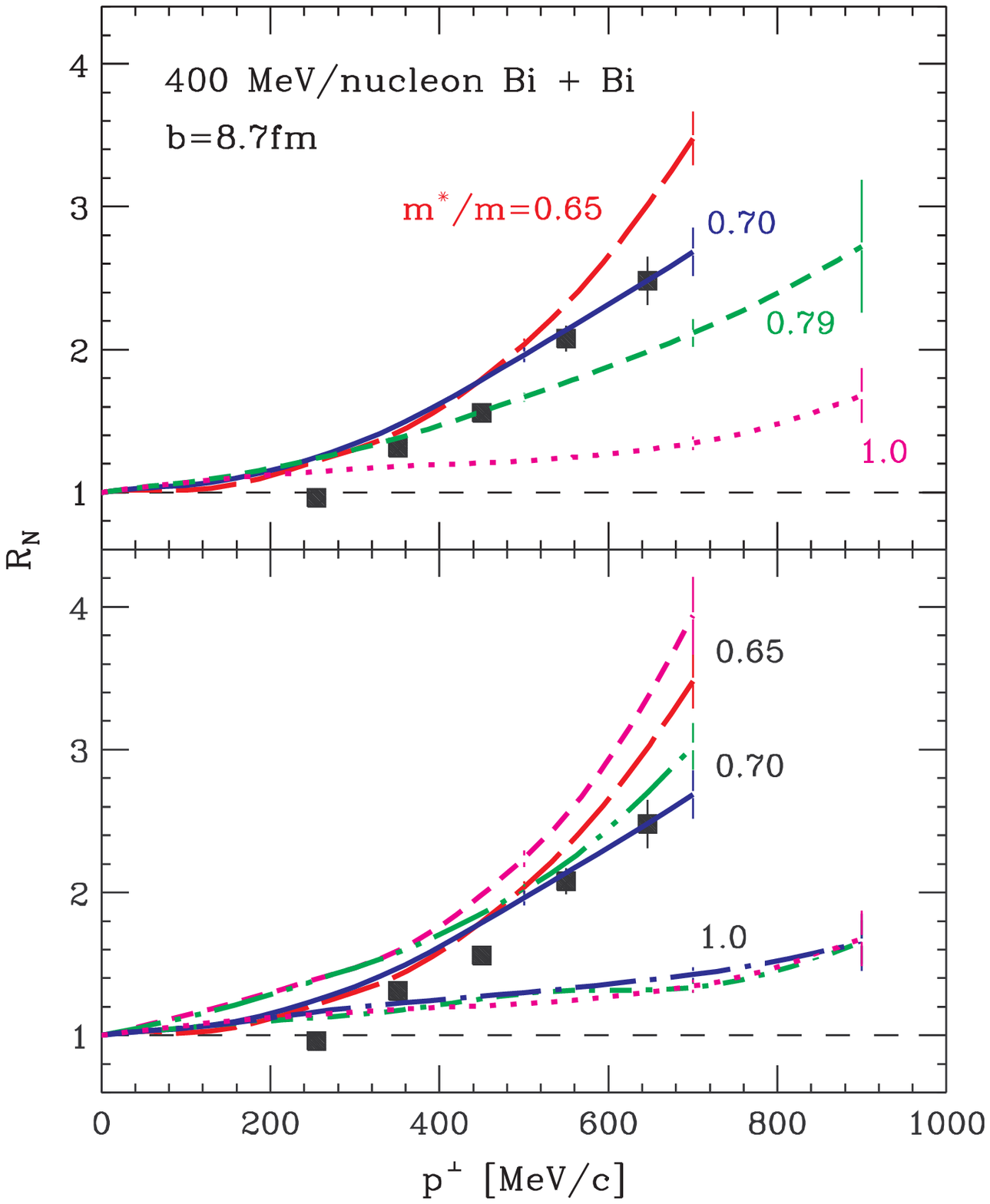}}
\vspace*{.1in}
\caption{
Measured~\protect\cite{bri96} (filled squares) and calculated
(lines) ratios
of out-of-plane to in-plane proton yields at midrapidity (0.35
$< y/y_{beam} <$ 0.65),  $R_N$ = $(N(90^\circ)$
+ $N(-90^\circ))/(N(0^\circ)$ + $N(180^\circ))$, as a~function
of transverse momentum, in 400~MeV/nucleon $^{209}$Bi +
$^{209}$Bi collisions at~$b =
8.7$~fm.  Both the measured and the calculated ratios are
computed from the
lowest Fourier coefficients of the proton distribution, $R_N
= (1 - v_2)/(1 + v_2)$.  The~lines in the top panel represent
results of the simulations done using in-medium cross sections
and those MFs
for which
the optical potentials are shown in Fig.~\protect\ref{upo1}.
The~numbers in the panels indicate the magnitude of effective
mass for the~MFs.  The~bottom panel shows the sensitivity of
the results to the variation of cross sections and
of incompressibility.
The~long-dashed, solid, and dotted lines repeat
respective
results from the top panel obtained with in-medium cross
sections and $K=210$~MeV.  The~long-dash-dotted and
short-dash-dotted lines represent additional
results obtained, respectively, using
the momentum-independent~MF corresponding to
$K=380$~MeV and using no MF at all.  The~short-dashed and
long-dash-double-dotted lines represent the additional results
obtained for free cross-sections and  MFs
corresponding to $m^*/m=0.65$ and to $m^*/m=0.70$,
respectively. }
\label{rnkt}
\end{figure}

To improve the reliability of transport reaction simulations
with momentum-dependent mean fields, and thus to enhance the
validity
of conclusions from comparisons to data, we~have generalized
the lattice Hamiltonian method~\cite{len89} to the case where
the reaction transport is formulated within the relativistic
Landau theory.   We have shown (Fig.~\ref{v2b}) that the
elliptic flow at midrapidity exhibits a~particularly strong
sensitivity to the mean-field momentum dependence in
midperipheral to
peripheral collisions.  A~relatively weak sensitivity was found
in these collisions to the incompressibility of nuclear matter.
An~additionally enhanced sensitivity to the mean-field momentum
dependence is exhibited by the elliptic flow of particles with
high transverse-momentum.  Variations in the mean field
associated with the change in the effective mass from $m^*/m =
0.79$ to $m^*/m=0.65$ change e.g.\ the high-$p^\perp$
anisotropy from $\sim 2$ to $\sim 3.5$ (Fig.~\ref{rnkt}).
Given that the
KaoS Collaboration has measured~\cite{bri96} the
$p^\perp$-dependence of elliptic anisotropies at three beam
energies and
at different impact parameters in Bi + Bi collisions, we went
on
to assess the mean-field momentum dependence from their data
and to assess the origin of the large high-$p^\perp$
sensitivity to the dependence.

\begin{figure}
\centerline{\includegraphics[angle=0,
width=.90\linewidth]{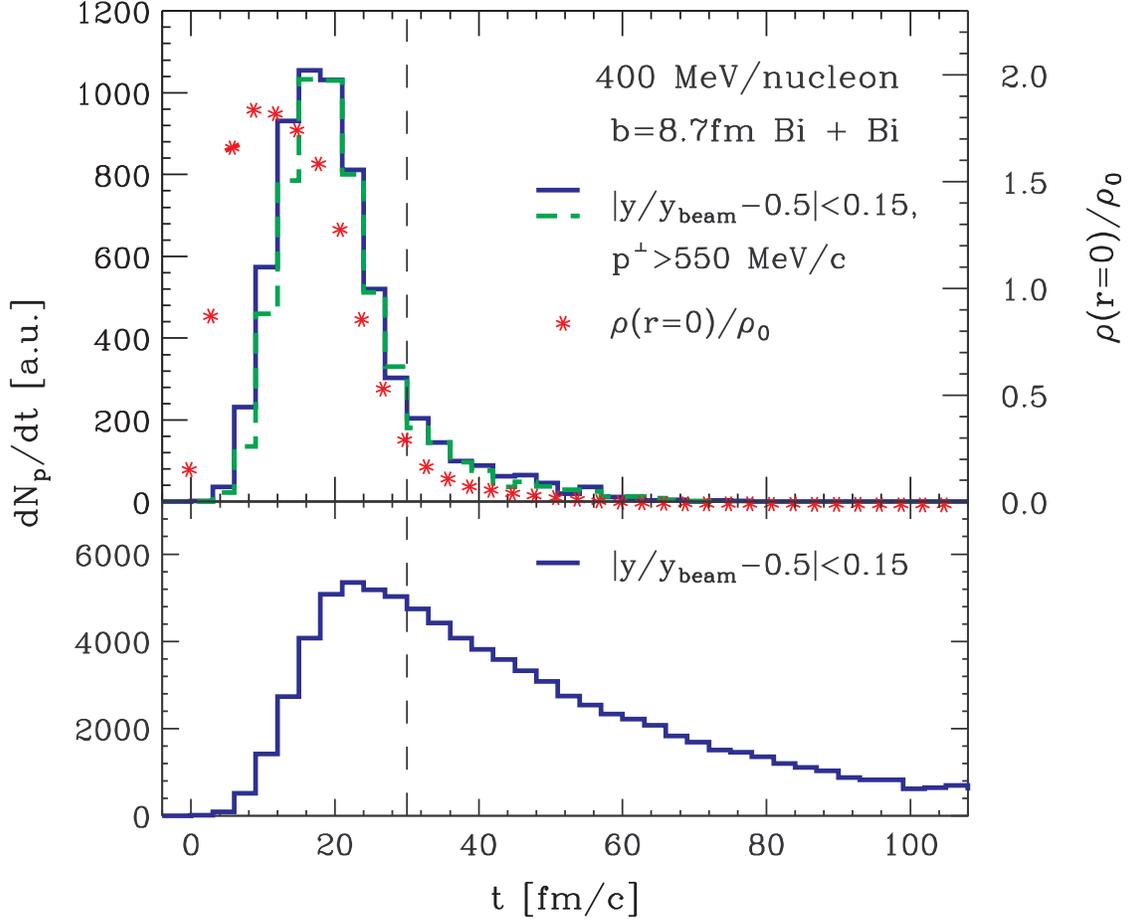}}
\vspace*{.1in}
\caption{
Time dependence of baryon density (stars in the top panel) at
the system center and time dependence of
midrapidity
emission-rate for high-momentum (lines in the top panel) and
all (bottom panel) protons, in Bi + Bi simulations
at 400~MeV/nucleon and
$b=8.7$~fm.  All results were obtained with the $K=210$~MeV
$m^*/m = 0.70$ MF, except for the emission rate for
the high-momentum protons represented by the thick dashed line,
obtained with the $K=380$~MeV momentum-independent~MF.
The~thin vertical dashed line at $t=30$~fm/c is drawn to guide
the eye.
}
\label{burst}
\end{figure}

We have shown in Fig.~\ref{burst} that the~high-density
stage in midperipheral collisions of heavy nuclei is
accompanied by a~burst of particles with high transverse
momenta.  These particles probe the high density matter in
a~quite direct and perturbative manner.  Thus, no
high-$p^\perp$ particles
are in practice emitted at any other time in a~reaction.
Remaining particles primarily continue along the beam
axis and leave the system at different times.
Looking at statistics in Fig.~\ref{burst}, it is seen that, even
amongst the emitted midrapidity particles, those with a~high
$p^\perp$ represent a~small fraction of all.  Thus,
the~high-$p^\perp$ particles
do not change the high-density environment as they leave.
The~elliptic
anisotropy of the high-$p^\perp$ particles is
sensitive to the sign and to the magnitude of the optical
potential felt by
these particles and, thus, to the~momentum dependence of the mean
field.
For a~repulsive potential, the~particles are speeded up as
they roll off the potential
step in a~transverse direction, escaping into the vacuum.

\begin{figure}
\centerline{\includegraphics[angle=0,
width=.75\linewidth]{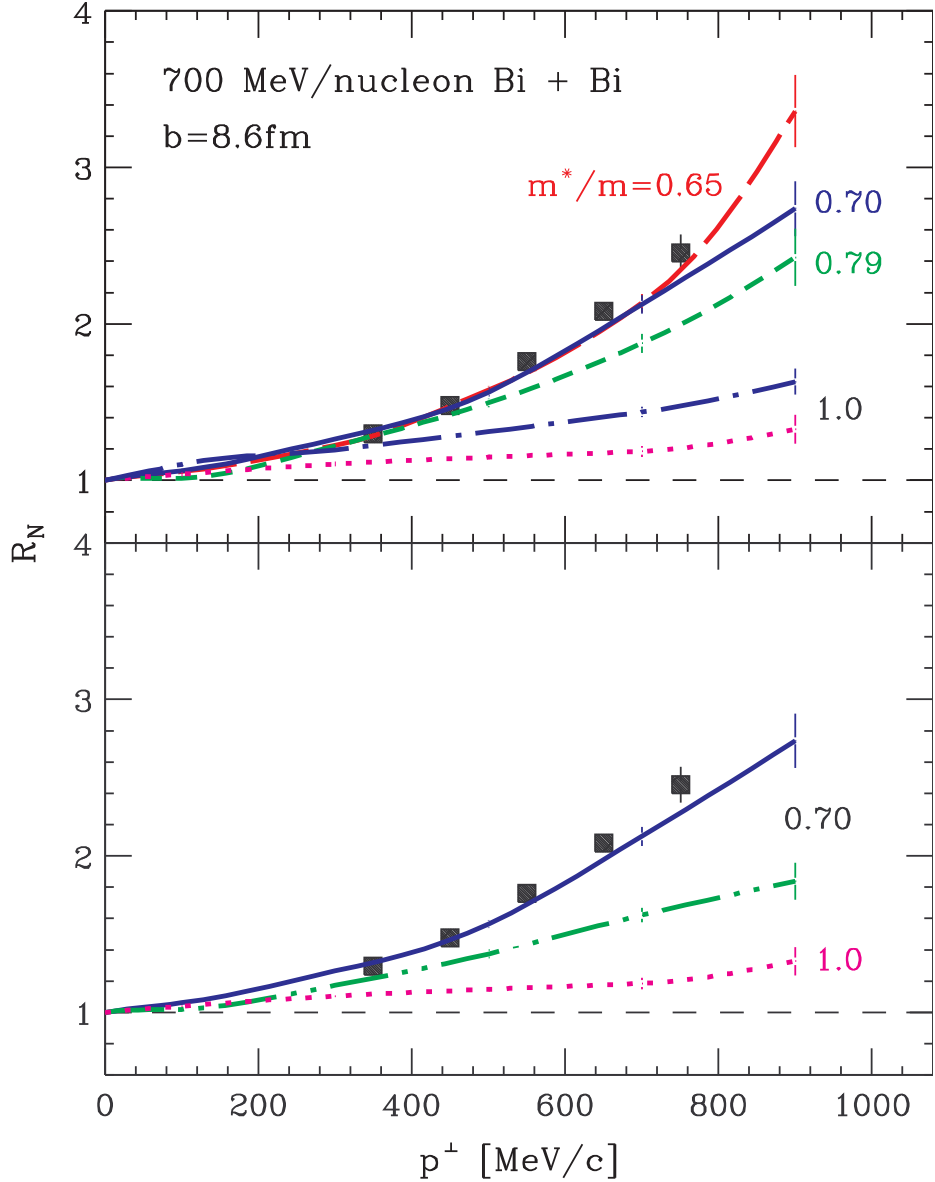}}
\vspace*{.1in}
\caption{
Measured~\protect\cite{bri96} (filled squares) and calculated
(lines) ratios
of out-of-plane to in-plane proton yields at midrapidity (0.35
$< y/y_{beam} <$ 0.65), as a~function
of transverse momentum, in 700~MeV/nucleon $^{209}$Bi +
$^{209}$Bi collisions at~$b =
8.6$~fm.  Both the measured and the calculated ratios are
computed from the
lowest Fourier coefficients of the proton distribution, $R_N
= (1 - v_2)/(1 + v_2)$.  The~numbers in the figure indicate the
values  of the effective mass in the ground state at Fermi
momentum for the calculations.  The~value of incompressibility
is $K=380$~MeV for the calculation represented by the
long-dash-dotted line in the top panel and $K=210$~MeV for
other calculations in that panel.
The~solid and dotted lines in the bottom panel repeat the
results from the top panel.  The~long-dash-double-dotted line
in the bottom panel represents the results of a~calculation
where
the momentum dependence of the MF at $\rho > \rho_0$ is made to
follow the dependence at~$\rho=\rho_0$, cf.\ the text.
}
\label{rnk7t}
\end{figure}

\begin{figure}
\centerline{\includegraphics[angle=0,
width=.75\linewidth]{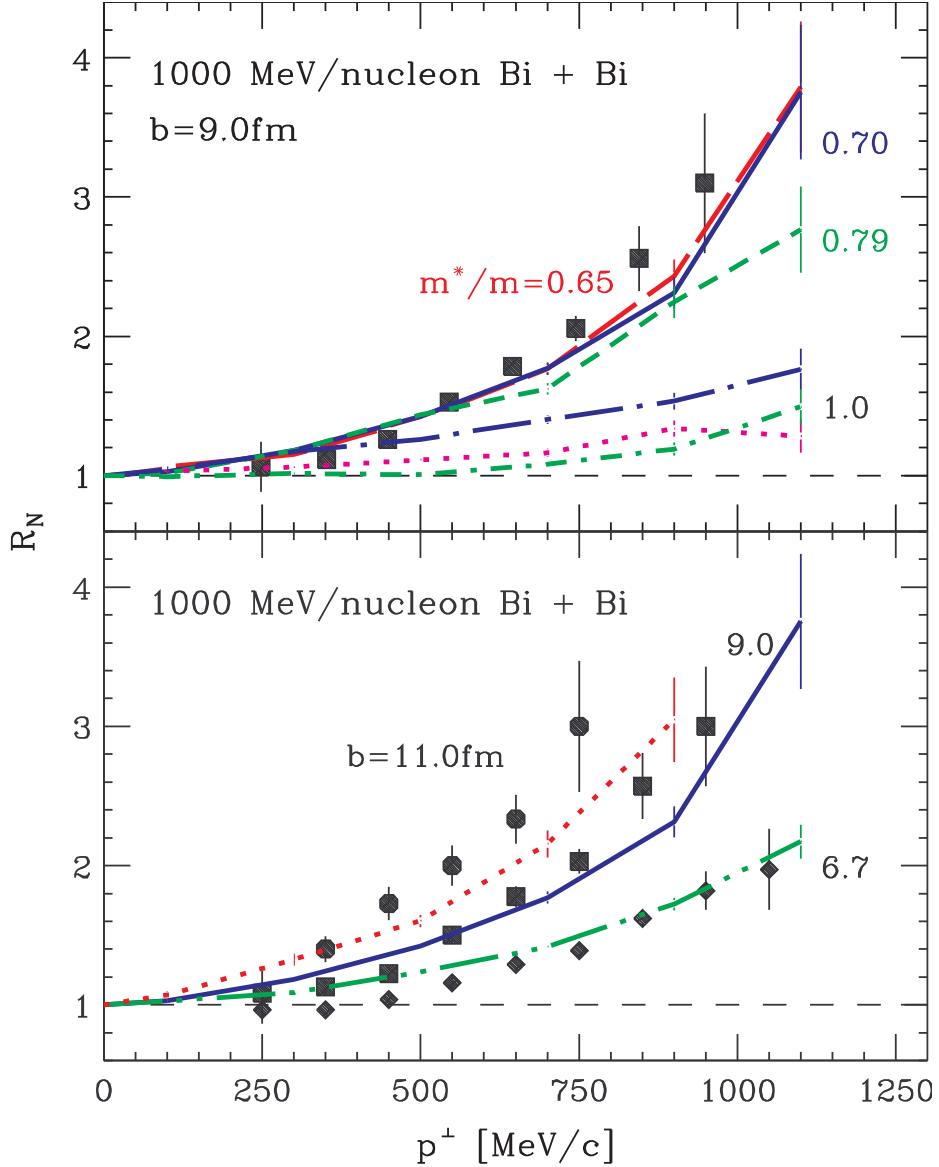}}
\vspace*{.1in}
\caption{
Measured~\protect\cite{bri96} (filled symbols) and calculated
(lines) ratios
of out-of-plane to in-plane proton yields at midrapidity (0.35
$< y/y_{beam} <$ 0.65), as a~function
of transverse momentum, in 1000~MeV/nucleon $^{209}$Bi +
$^{209}$Bi collisions.
Both the measured and calculated ratios are computed
from the
lowest Fourier coefficients of the proton distribution, $R_N
= (1 - v_2)/(1 + v_2)$.  The~top panel shows the ratios for
collisions at~$b = 9.0$~fm.  The~long-dashed, solid, and
short-dashed lines represent calculations for the
momentum-dependent~MFs corresponding to $K=210$~MeV and $m^*/m
=$ 0.65 (first of the sets in Table~\protect\ref{tabmo}),
$m^*/m = 0.70$,
and~0.79, respectively.  The~long-dash-dotted, dotted, and
short-dash-dotted lines represent two calculations with
momentum-independent~MFs corresponding to $K=380$~MeV,
and $K=210~MeV$, and one with no MF at all, respectively.
The~numbers in the panel indicate values of the effective mass
for the calculations.
The~bottom panel shows ratios for
collisions at impact parameters $b = 11.0$~fm (filled circles
and dotted line), 9.0~fm (filled squares and solid line), and
6.7~fm (filled diamonds and long-dash-double-dot line).
The~calculations for the bottom panel were carried out using
the~MF
characterized by $K=210$~MeV and $m^*/m = 0.70$.  The~numbers
in the bottom panel indicate the values of impact
parameter.
}
\label{rnk10t}
\end{figure}

The~large anisotropies observed by the KaoS Collaboration
at high momenta in the midperipheral collisions at
400--1000~MeV/nucleon (Figs.~\ref{rnkt}--\ref{rnk10t})
can only be explained in the transport simulations when
assuming optical potentials
with a~momentum dependence that strengthens as the nuclear
density increases beyond the normal density.  For momenta where
the large anisotropies are observed, the~potential must be strongly
repulsive; the~changes in the potential as low as 10~MeV yield
observable changes in the anisotropy.  The~particular data
constraint the potential in simulations at
densities from within the region of $\sim (1.4-2.3) \, \rho_0$ and
at momenta which correspond to free-space kinetic energies of
$\sim (100-400)$~MeV.

We have compared optical potentials from the MF parametrizations
in the transport simulations to the potentials from microscopic
nuclear-matter calculations.   Potentials
in the simulations allowed for a~different quality of
description of the measured anisotropies
depending on beam energy and
thus densities reached.  We found, in the potential
comparisons, that two
of the microscopic potentials, BBG~\cite{bal89,ins94} and AV14
+ UVII~\cite{wir88}, were unacceptably weak at supranormal
densities and probed energies.  On the other hand,
the~potentials from DBHF calculations~\cite{bro90,lee97} and
the UV14 + TNI potential from the variational
calculations~\cite{wir88} turned out to be surprisingly close,
in their momentum and density dependence within the accessed
region of dependence in reactions, to the~parametrized
potentials giving an~acceptable description of the data.

\begin{figure}

\centerline{\includegraphics[angle=0,
width=.70\linewidth]{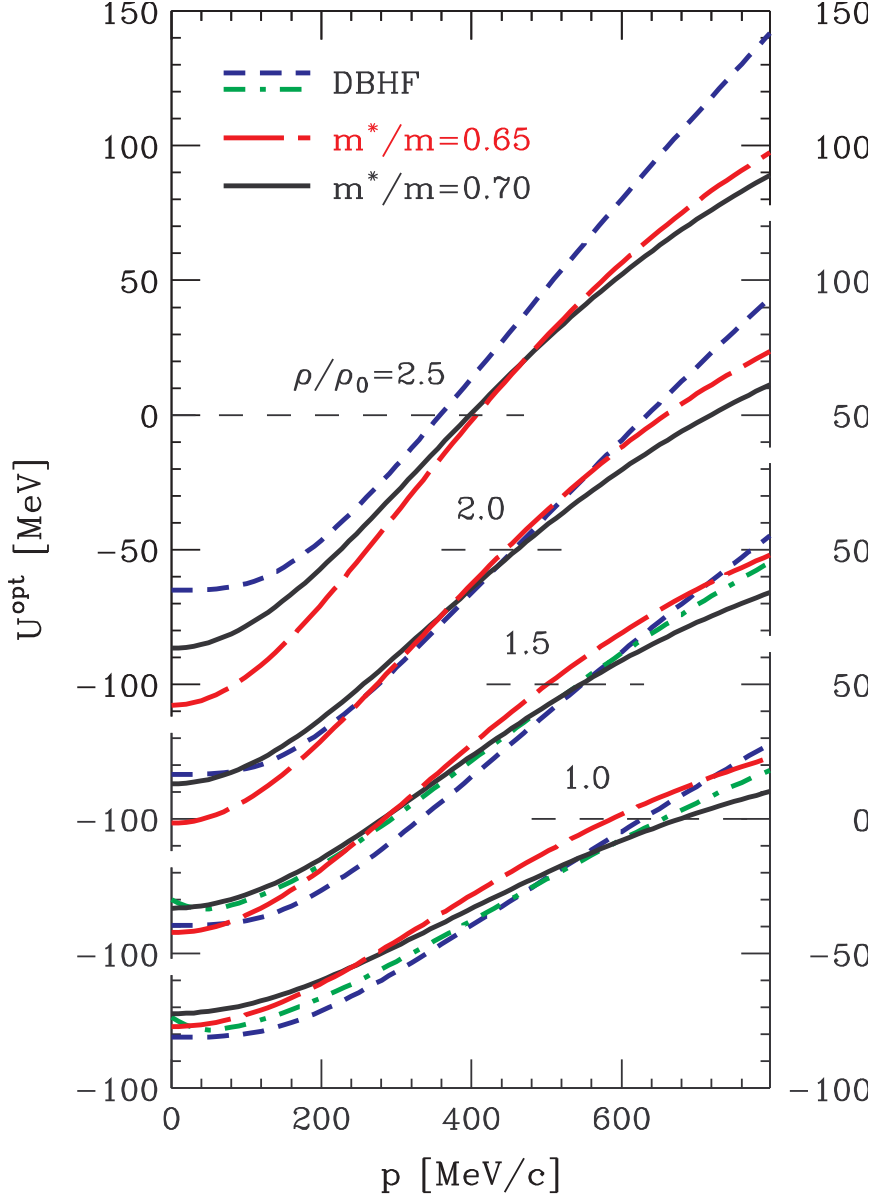}}
\vspace*{.1in}
\caption{
Optical potential in nuclear matter as
a~function of nucleon momentum, at different densities, from
the DBHF
calculations with Bonn-A interaction~\protect\cite{bro90} and
in our parametrizations.
The~short-dashed and short-dash-dotted lines
represent the DBHF
potentials obtained assuming, respectively,
momentum-independent vector and
scalar~MFs~\protect\cite{lim93,bro90} and parametrized
momentum-dependent~MFs~\protect\cite{lee97}.
The~solid and long-dashed lines represent the optical
potentials for our $m^*/m=0.70$ and first $m^*/m=0.65$
$K=210$~MeV MF parametrizations, respectively.
The~numbers
in the figure indicate the values of density in units
of the normal density $\rho_0=0.16$~fm$^{-3}$.
The~thin horizontal dashed lines indicate the zero
value for the potential.
}
\label{ugqli}
\end{figure}

\begin{figure}

\centerline{\includegraphics[angle=0,
width=.70\linewidth]{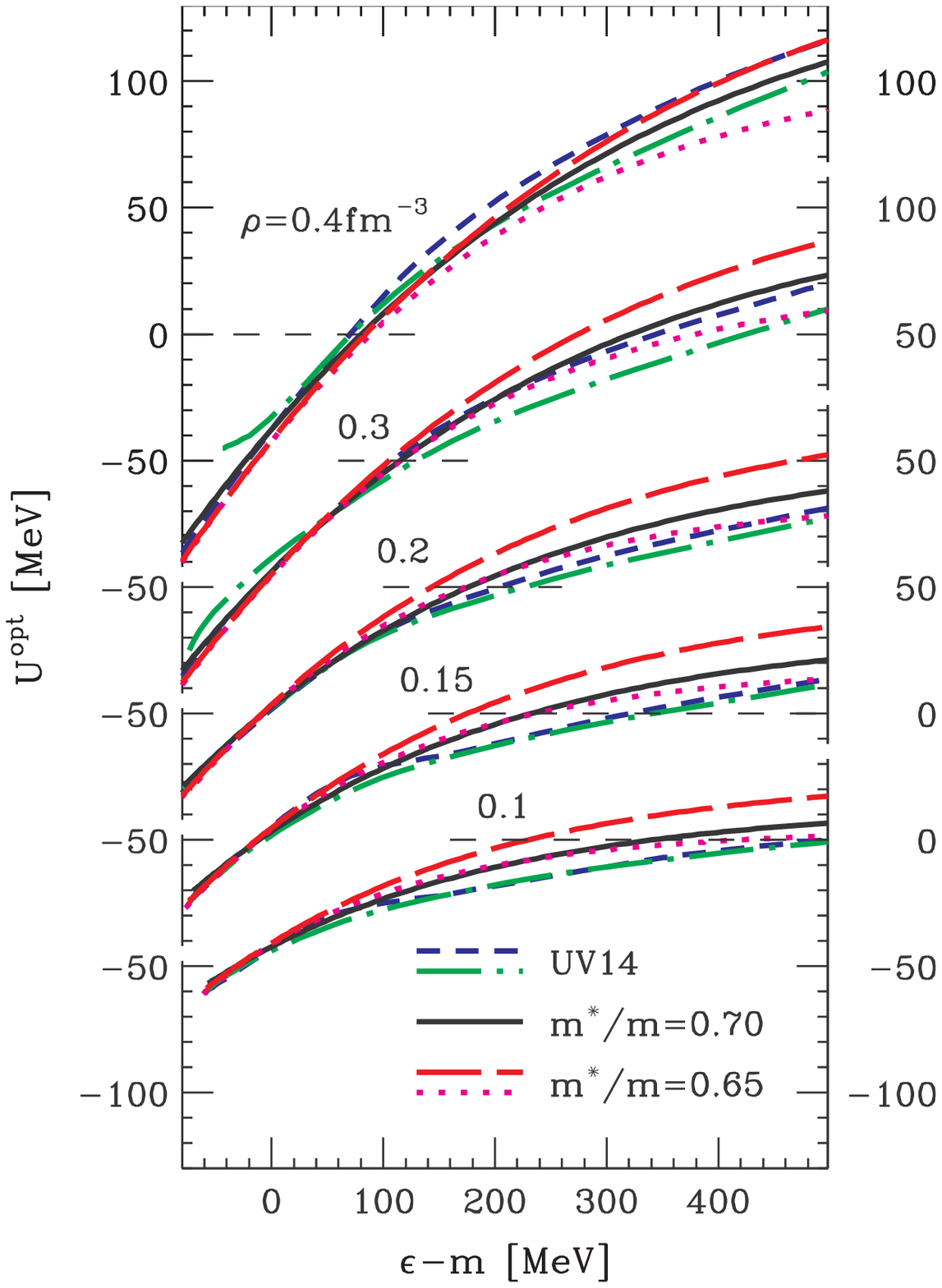}}
\vspace*{.1in}
\caption{
Optical potential in nuclear matter as
a~function of nucleon energy, at different densities, from
the variational
calculations with UV14 + TNI and UV14 + UVII
interactions~\protect\cite{wir88} and in our parametrizations.
The~short-dashed and long-dash-dotted lines represent the UV14
+ TNI and UV14 + UVII potentials, respectively.
The~solid, long-dashed, and dotted lines represent,
respectively,
the~potentials for our $m^*/m=0.70$, and first and second
$m^*/m=0.65$ $K=210$~MeV MF parametrizations.
The~numbers
in the figure indicate the values of density in units
of fm$^{-3}$.
The~thin horizontal dashed lines indicate the zero
value for the potential.
}
\label{uwirb}
\end{figure}

We hope that present and possible future results from
flow in peripheral collisions may play a~similar role to the
results from nucleon-nucleus scattering in constrainting
microscopic theory, but in this case at supranormal densities.
A~combination of microscopic theories with two sets of
constraints can make extrapolations to the regions of
uncertainty, such as~low momenta in matter at high density,
more trustworthy.
Incidentally, in~Figs.~\ref{ugqli} and \ref{uwirb} there is
quite a~degree of convergence in the
low-momentum region between the potentials which agree in the
high-momentum region.

With regard to possible future investigations of flow,
we~encountered here ambiguities at low transverse momenta
of emitted particles that might become clarified.  The~study
should be extended to high transverse momenta, which requires
increasing statistics within simulations.  Moreover,
investigations could be extended down~\cite{tsa96} and
up~\cite{pin99} in the beam energy.
The~first-order flow, when analyzed in a~similar manner to the
second-order,
is likely to contain a~comparable amount of
information~\cite{pan93,zha94,lik96,pak96,sof95}.

Better constraints on the momentum dependence of the optical
potentials
should make the determinations of the nuclear equation of state
more reliable~\cite{pan93,zha94}.  These searches should
rather be carried at
the more central impact parameters (Fig.~\ref{v2b}) where the
matter is better equilibrated and reaches higher densities.
One recent example, illustrating the importance of clarifying
the momentum dependence in that context, is~Ref.~\cite{sah99}
where, in the simulations for AGS energies, the~baryon optical
potential is put to zero above a~cut-off
momentum.  This should lead to a~discontinuity in the potential
growing with density and likely generate effects similar to
a~phase transition that one looks for in that energy
regime~\cite{pin99}.
Notably, at high incident energies, even in midperipheral
collisions, the~matter is dominated by resonances, so
investigations of proton flow are likely to mix information on
the potential felt by nucleons with that felt
by baryon resonances.

\acknowledgements

This paper was possible due to a~progress in collaborations on
related projects with a~number of colleagues, including
N.~Ajitanand, P.~Bo\.zek, P.-B.~Gossiaux, and R.~Lacey.
Y.~Leifels helped in understanding a~set of data pertinent to
the paper.  Some of the writing was carried out during a~stay
at the Institute
for Nuclear Theory in Seattle.  This work was partially
supported by the National Science Foundation
under Grant PHY-9605207 and by the U.S.\ Department of Energy.

\newpage
\begin{table}
\caption{Parameter values for the momentum-dependent mean
fields.  The momentum $p_0$ is the one for which the optical
potential vanishes at the normal density, $U_{opt} (p_0,
\rho_0)
= 0$, while $U_{opt}^\infty$ is the asymptotic value of the
potential at $\rho_0$ as $p \rightarrow \infty$.
The~annotated sets have been used
in~\protect\cite{dan98}.}
\begin{center}
\begin{tabular}{ c c c c c c c c c c}
{$a$} & $b$ & $\nu$ & $c$ & $\lambda$ &
$m^*/m$ & $p_0$ & $U_{opt}^\infty$ & K & Ref. \\
{(MeV)} & (MeV) &  &  & & & (MeV) & (MeV) &
(MeV) \\ \hline 185.47& 36.291 &
1.5391 & 0.83889 & 1.0890 & 0.65 & 585 & 55 & 210 & \protect\cite{dan98}  \\
185.56 & 32.139 & 1.5706 & 0.96131 & 2.1376 & 0.65 & 680 & 23
& 210 \\
209.79 & 69.757 & 1.4623 & 0.64570 & 0.95460 & 0.70 & 680 & 40
& 210 \\
214.10 & 95.004 & 1.4733 & 0.37948 & 0.55394 & 0.79 & 900 & 25
& 210 \\
123.62 & 14.653 & 2.8906 & 0.83578 &
1.0739 & 0.65 & 580 & 56 & 380 & \protect\cite{dan98}  \\
128.22 & 22.602 & 2.5873 & 0.64570 & 0.95460 & 0.70 & 685 & 39
& 380 \\
\end{tabular}
\end{center}
\label{tabmo}
\end{table}

\begin{table}
\caption{Accuracy of energy conservation and supplementary
information
for simulations of $b=0$ Bi + Bi collisions at different
beam energies $T_{beam}/A$, using the
momentum-dependent MF given by the first of the value
sets listed in Table~\protect\ref{tabmo}.  The~first number for
$\Delta_{max}
(\overline{E}/A)$ is the maximal deviation from the initial cm
energy for a~simulation with
the MF only; the second number is for a~simulation with
the MF and collisions.  The~simulations utilized
$\Delta l_1 = \Delta l_2 = 0.92$~fm and ${\cal N} = 170$.
The~accuracy improves for lighter nuclei and
momentum-independent fields (5~times as a~rule, in the
latter case). }
\begin{center}
\begin{tabular}{l  c c c}
$T_{beam}/A$  [GeV/nucleon] & 0.040 & 0.400 & 10.7 \\
\hline
$\Delta l_3$  [fm]  & 0.92 & 0.92 & 0.40 \\
$\Delta t$  [fm/c] & 0.50 & 0.30 & 0.080 \\
$t_{max}$  [fm/c] & 300 & 110 & 40 \\
$\Delta_{max} (\overline{E}/A)$  [MeV/nucleon]
& 0.9/0.6 & 3.9/3.5 &  29/55 \\
\end{tabular}
\end{center}
\label{enco}

\end{table}

\newpage

\newpage

\newpage

\newpage
\begin{references}

\bibitem{men71}
J.\ J.\ H.\ Menet, E.\ E.\ Gross, J.\ J.\ Malinify, and A.\
Zucker, \prc{4}, 1114 (1971).

\bibitem{sch82}
P.\ Schwandt, \prc{26}, 55 (1982); AIP Conf.\ Proc. 97, 89
(1983).


\bibitem{ham90}
S.\ Hama \etal, \prc{41}, 2737 (1990).

\bibitem{kle94}
M.\ Kleinmann, R.\ Fritz, H.\ M\"uther, and A.\ Ramos,
\npa{579}, 85 (1994).

\bibitem{wir88}
R.\ B.\ Wiringa, \prc{38}, 2967 (1988).

\bibitem{bal89}
M.\ Baldo, I.\ Bombaci, G.\ Giansiracusa, and U.\ Lombardo,
\prc{40}, R491 (1989).

\bibitem{lim93}
G.\ Q.\ Li and R.\ Machleit, \prc{48}, 2707 (1993).

\bibitem{ins94}
A.\ Insolia, U.\ Lombardo, N.\ G.\ Sandulescu, and A.\
Bonasera, \plb{334}, 12 (1994).

\bibitem{lee97}
C.-H.\ Lee, T.\ T.\ S.\ Kuo, G.\ Q.\ Li, and G.\ E.\ Brown,
\plb{412}, 235 (1997).

\bibitem{pan93}
Q.\ Pan and \pdd, \prl{70}, 2062, 3523 (1993).

\bibitem{zha94}
J.~Zhang, S.~Das Gupta, and C.~Gale, \prc{50},
      1617, (1994).



\bibitem{pak96}
R.\ Pak \etal, \prc{53}, R1469 (1996).

\bibitem{hun96}
M.\ J.\ Huang \etal, \prl{77}, 3739 (1996).


\bibitem{bay76}
G.\ Baym and S.\ U.\ Chin, Nucl.\ Phys.\ A262, 527 (1976).

\bibitem{dan98}
\pdd\ \etal, \prl{81}, 2438 (1998).

\bibitem{hol77}
G.~Holzwarth, \plb{66}, 29 (1977).

\bibitem{len89}
R.\ J.\ Lenk and V.\ R.\ Padharipande, \prc{39}, 2242 (1989).


\bibitem{rin80} P.\ Ring and P.\ Schuck, {\em The Nuclear
Many-Body Problem} (Springer-Verlag, New York, 1980).

\bibitem{bec69}
F.~D.~Becchetti and G.~R.~Greenlees, Phys.\ Rev.\ 182, 1190
(1969).

\bibitem{mye98}
W.~D.~Myers and W.~\'{S}wiatecki, \prc{57}, 3020 (1998).

\bibitem{hom99}
A.~Hombach, W.~Cassing, S.~Teis, and U.~Mosel, Eur.\ Phys.\ J.\
A 5, 157 (1999).

\bibitem{ber88}
G.\ F.\ Bertsch and S.\ Das Gupta, Phys.\ Rep.\ 160,
189 (1988).

\bibitem{cse92}
L.~P.~Csernai, G.~Fai, C.~Gale, and E.~Osnes, Phys.\ Rev.\
C 46, 736 (1992).

\bibitem{jam89}
M.~Jaminon and C.~Mahaux, \prc{40}, 354 (1989).


\bibitem{jag74}
C.\ W.\ de Jager, H.\ de Vries, and C.\ de Vries,
At.\ Data Nucl.\ Data Tables {14}, 479 (1974).

\bibitem{fel91}
H.~Feldmeier and J.~Lindner, \zpa{341}, 83 (1991).

\bibitem{har87}
B.~ter Haar and R.~Malfliet, Phys.\ Rep.\ 149, 207 (1987).


\bibitem{per76}
C.~M.~Perey and F.~G.~Perey, At.\ Data Nucl.\ Data Tables
17, 1 (1976).

\bibitem{bri77}
F.\ A.\ Brieva and J.\ R.\ Rook, \npa{291}, 299 (1977).

\bibitem{fri81}
B.\ Friedman and V.\ R.\ Pandharipande, \plb{100}, 205 (1981).

\bibitem{per99}
D.~Persram and C.~Gale, nucl-th/9901019.

\bibitem{fel92}
H.\ Feldmeier and \pdd, National Superconducting Cyclotron
Laboratory Report MSUCL-833 (1992), unpublished.

\bibitem{dic98}
W.\ H.\ Dickhoff, \prc{55}, 2807 (1998).

\bibitem{boz99}
P.\ Bo\.zek, \prc{59}, 2619 (1999).

\bibitem{dan99}
in preparation.

\bibitem{dan91}
     P.~Danielewicz and G.\ F.\ Bertsch, Nucl.\ Phys.\ A 533,
712 (1991).

\bibitem{lan93}
A.\ Lang \etal, J.\ Comp.\ Phys.\ 106, 391 (1993).

\bibitem{gal90}
C.\ Gale \etal, \prc{41}, 1545 (1990).

\bibitem{lam94}
D.\ Lambrecht \etal, \zpa{350}, 115 (1994).

\bibitem{bri96}
D.\ Brill \etal, \zpa{355}, 61 (1996).

\bibitem{dan95}
P.~Danielewicz, \prc{51}, 716 (1995).

\bibitem{and99}
A.~Andronic, private communiction, 1999.

\bibitem{bro90}
R.~Brockmann and R.~Machleit, \prc{42}, 1965 (1990).

\bibitem{fuc96}
C.~Fuchs, T.~Gaitanos, and H.~H.~Wolter, \plb{381}, 23 (1996);
L.~Sehn and H.~H.~Wolter, \npa{601}, 473 (1996).


\bibitem{tsa96}
M.~B.~Tsang \etal, \prc{53}, 1959 (1996).

\bibitem{pin99}
C.~Pinkenburg \etal, \prl{83}, 1295 (1999).

\bibitem{lik96}
B.-A.~Li, C.~M.~Ko, and G.~Q.~Li, \prc{54}, 844 (1996).

\bibitem{sof95}
S.~Soff \etal, \prc{51}, 3320 (1995).


\bibitem{sah99}
P.~K.~Sahu, W.~Cassing, U.~Mosel, and A.~Ohnishi, nucl-th/9907002.


\end{references}
\end{document}